\documentclass[aps,superscriptaddress,twocolumn,prapplied]{revtex4-2}
\usepackage{lineno}
\usepackage{graphicx}
\usepackage{dcolumn}
\usepackage{bm}
\usepackage[utf8]{inputenc}
\usepackage[T1]{fontenc}
\usepackage{mathptmx}
\usepackage{etoolbox}
\usepackage{upgreek}
\usepackage{mathrsfs}
\usepackage{hyperref} %
\usepackage{amsmath}
\usepackage{float}

\begin{document}


\title{Accelerated Bayesian optimization in deep cooling atoms}


\author{Xiaoxiao Ma}
\thanks{These authors contributed equally to this work.}
\affiliation{College of Intelligence Science and Technology, National University of Defense Technology, Changsha 410073, China}
\affiliation{Interdisciplinary Center for Quantum Information, National University of Defense Technology, Changsha 410073, China}

\author{Changwen Liang}
\thanks{These authors contributed equally to this work.}
\affiliation{College of Intelligence Science and Technology, National University of Defense Technology, Changsha 410073, China}
\affiliation{Interdisciplinary Center for Quantum Information, National University of Defense Technology, Changsha 410073, China}

\author{Rong Sha}
\thanks{These authors contributed equally to this work.}
\affiliation{College of Intelligence Science and Technology, National University of Defense Technology, Changsha 410073, China}
\affiliation{Interdisciplinary Center for Quantum Information, National University of Defense Technology, Changsha 410073, China}

\author{Chao Zhou}
\affiliation{College of Intelligence Science and Technology, National University of Defense Technology, Changsha 410073, China}
\affiliation{Interdisciplinary Center for Quantum Information, National University of Defense Technology, Changsha 410073, China}

\author{Qixue Li}
\affiliation{College of Intelligence Science and Technology, National University of Defense Technology, Changsha 410073, China}
\affiliation{Interdisciplinary Center for Quantum Information, National University of Defense Technology, Changsha 410073, China}

\author{Guochao Wang}
\affiliation{College of Intelligence Science and Technology, National University of Defense Technology, Changsha 410073, China}
\affiliation{Interdisciplinary Center for Quantum Information, National University of Defense Technology, Changsha 410073, China}

\author{Jixun Liu}
\affiliation{College of Intelligence Science and Technology, National University of Defense Technology, Changsha 410073, China}
\affiliation{Interdisciplinary Center for Quantum Information, National University of Defense Technology, Changsha 410073, China}

\author{Shuhua Yan}
\affiliation{College of Intelligence Science and Technology, National University of Defense Technology, Changsha 410073, China}
\affiliation{Interdisciplinary Center for Quantum Information, National University of Defense Technology, Changsha 410073, China}

\author{Jun Yang}
\affiliation{College of Intelligence Science and Technology, National University of Defense Technology, Changsha 410073, China}
\affiliation{Interdisciplinary Center for Quantum Information, National University of Defense Technology, Changsha 410073, China}

\author{Lingxiao Zhu}
\email[contact author: ]{zhulingxiao31@163.com}
\affiliation{College of Intelligence Science and Technology, National University of Defense Technology, Changsha 410073, China}
\affiliation{Interdisciplinary Center for Quantum Information, National University of Defense Technology, Changsha 410073, China}

\date{\today}

\begin{abstract}
Laser cooling, which cools atomic and molecular gases to near absolute zero, is the crucial initial step for nearly all atomic gas experiments. However, fast achievement of numerous sub-$\upmu$K cold atoms is challenging. To resolve the issue, we propose and experimentally validate an intelligent polarization gradient cooling approach enhanced by optical lattice, utilizing Maximum Hypersphere Compensation Sampling Bayesian Optimization (MHCS-BO). MHCS-BO demonstrates a twofold increase in optimization efficiency and superior prediction accuracy compared to conventional Bayesian optimization. Finally, approximate 10$^8$ cold atoms at a temperature of 0.4$\pm$0.2 $\upmu$K can be achieved given the optimal parameters within 15 minutes. Our work provides an intelligent protocol, which can be generalized to other high-dimension parameter optimization problems, and paves way for preparation of ultracold atom in quantum experiments.
\end{abstract}


\maketitle

\section{introduction}
Atomic cooling plays a crucial role in various quantum systems, including quantum precision measurement \cite{peters1999,jones2006may,safronova2018jun,cimini2023feb}, quantum simulation \cite{ciavarella2023,ott2021,halimeh2022}, and quantum information \cite{cao2024oct,mol2023,vovcenko2021}. Atomic cooling methods include magneto-optical trap (MOT) cooling \cite{kasevich1991}, polarization gradient cooling (PGC) \cite{ejtemaee2017jul}, evaporation cooling \cite{wang2021}, Raman sideband cooling \cite{zohar2022} and so on. Among these, PGC stands out as a cheap approach that could obtain lower atomic temperature simply by designing timing sequence without the need for extra hardware. Researches indicate that in quantum metrology based on cold atom interferometry, lower atom temperature could improve interference fringe contrast and measurement sensitivity \cite{debs2011sep,louchet-chauvet2011}. While conventional PGC is fundamentally limited to $\upmu$K regime, evaporative cooling can prepare Bose-einstein condensation (BEC) at nK level. However, the generation of BEC usually requires demanding hardware (e.g., high-power optical trap) and exhibits low efficiency \cite{anderson1995jul,vanderstam2007jan,xu2024jun}, which is a critical limitation incompatible with the demand for high measurement frequency in quantum metrology. Thus, the efficient preparation of sub-$\upmu$K cold atoms hold critical importance for enhancing atom interferometer performance and enabling engineering applications. With the aid of optical lattice to adiabatic cooling, the atomic temperature could be further reduced \cite{wei2018jun}. However, the process of preparing cold atoms is quite complex, highly nonlinear, and susceptible to environmental influences. Conventional manual optimization results in low accuracy and efficiency, making it difficult to control the cooling process more finely due to the complexities associated of high-dimension optimization. It is also particularly challenging to ensure a lower temperature while maintaining a higher atom number.

Recently, artificial intelligence (AI) has been applied to solve optimization problems involving quantum systems \cite{wigley2016,carleo2017,torlai2018}. Evolutionary algorithm such as genetic algorithm \cite{rohringer2008}, differential evolution \cite{palittapongarnpim2017}, particle swarm optimization \cite{che2022jul_075211}, can search global optimal parameters. However, these algorithms, usually depending on large population scale and iteration steps, work inefficiently in scenarios with high experimental cost. Neural networks (NNs) such as multi-layer perceptron \cite{liu2021}, deep learning \cite{tranter2018,vendeiro2022dec}, reinforcement learning \cite{ledesma2023,reinschmidt2024oct} are applicable to high-dimensional multi-objective optimization problems. Similar to Evolutionary algorithm, NNs also require a large amount of experimental data to train networks. The accuracy of NNs extremely requires robust system and high-quality datasets. Fortunately, Bayesian optimization (BO) \cite{barker2020mar,xu2021} can efficiently optimize parameters especially in complex systems with high experimental cost. The acquisition function, e.g., expected improvement (EI), probability of improvement and upper confidence bound, could obtain good optimization effects in the majority of scenarios. Nevertheless, the traditional acquisition functions always leave large non-sampling regions in parameter space, compromising the predictive performance of the model in high dimension.

\begin{figure*}
	\centering
	\includegraphics{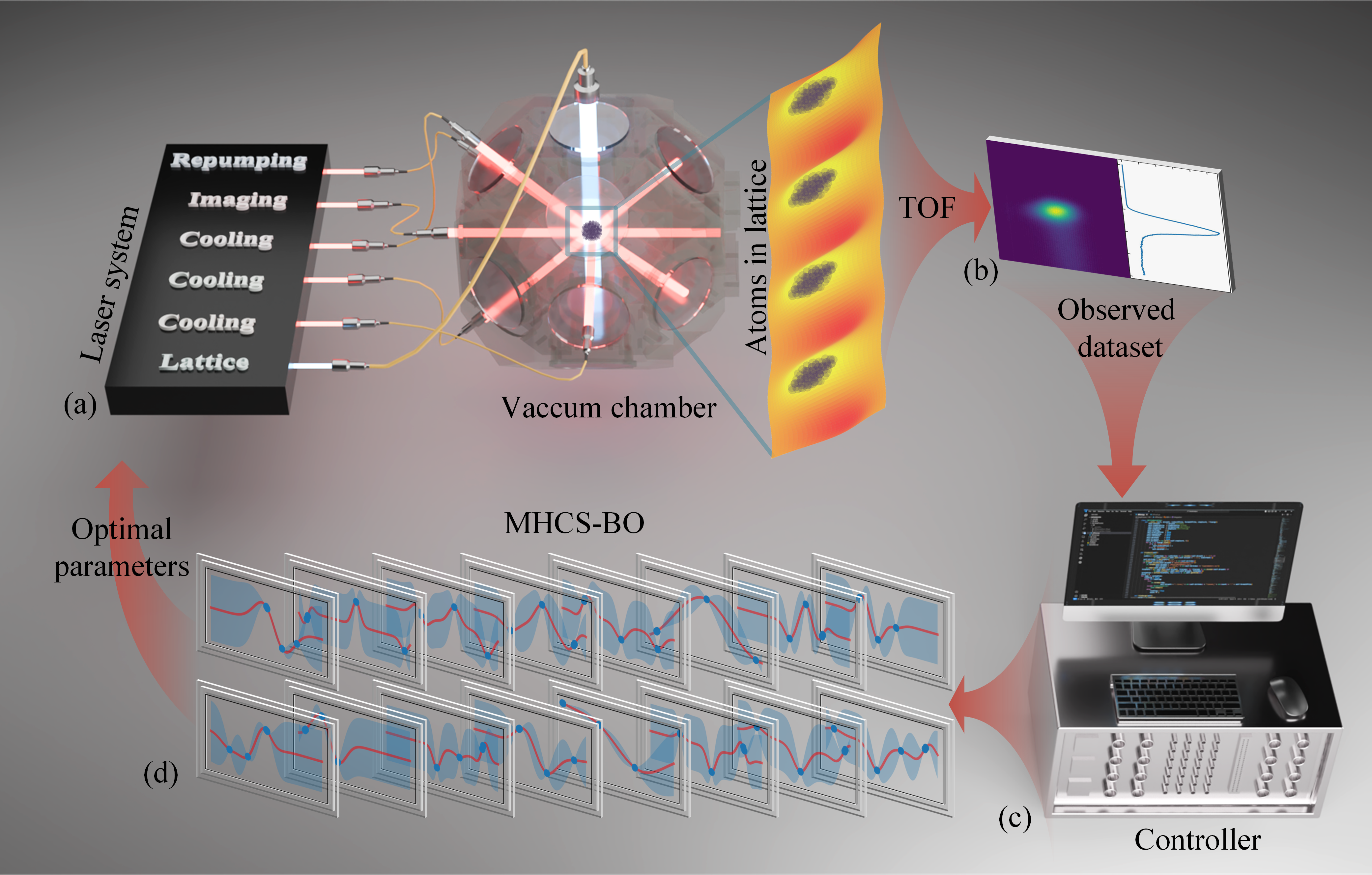}
	\caption{Experiment diagram. (a) The atoms are trapped in a 3D MOT, where the PGC assisted by optical lattice can be achieved through controlling the experimental parameters such as magnetic field, laser power and detuning. The 1-dimension optical lattice is formed by two red detuning laser beams, with intensity distribution following Gaussian and harmonic profile along the radial and axial directions respectively. The atoms are localized at the peak of lattice light. (b) After being released in the vacuum chamber, the atom cloud would reach towards detection zone, generating a TOF signal that can be utilized to calculate the objective function. Each pairwise objective function and corresponding parameters constitutes an observed data point which will be employed for training the Bayesian model. A host computer executes the MHCS-BO algorithm with observed dataset (d), predicting the optimal parameters which are utilized in the laser system through the controller (c).}
	\label{fig:sb}
\end{figure*}
In our previous work~\cite{liang2024nov}, we utilized reinforcement learning (RL) to optimize PGC sequence, demonstrating superior cooling performance over manual tuning while validating the feasibility of intelligent fine control over PGC process. In this study, we implement one-dimensional deep cooling along the vertical direction assisted by optical lattice to obtained sub-$\upmu$K cold atoms. The increased system complexity and strong parameter coupling increase the cost of objective function, which would lead to lower efficiency using RL framework. We utilize Bayesian Optimization (BO) to address this challenge. As an efficient global optimization framework well-suited for expensive objective functions, BO demonstrates superior optimization efficiency compared to evolutionary algorithms and neural networks \cite{barker2020mar,anton2024jun}. Furthermore, we make improvements to the traditional BO and proposed Maximum Hypersphere Compensation Sampling Bayesian Optimization (MHCS-BO) to perform stepwise decoupled optimization of the cooling process. It is illustrated that MHCS-BO has twice optimization efficiency and higher prediction accuracy than traditional BO. As a result, lower atom temperature and more sufficient atoms can be obtained. In the experiment, the cooling process is divided into 10 segments to be finely controlled, and thus a total of 16 experimental parameters is optimized in a decoupled manner by MHCS-BO. Approximate 10$^8$ cold atoms at a temperature of 0.4±0.2 $\upmu$K can be obtained with 15 minutes of MHCS-BO.

\section{Experiment setup}
As described in our previous work \cite{liang2024nov}, the experiment setup is mainly composed of 3D-MOT, laser systems (see Appendix \ref{app:laser}), vacuum chamber, and controller (FIG.~\ref{fig:sb}). The cooling beam has a radius of about 26 mm, a power of about 14 mW, and is red-detuned by 15 MHz from $|F=2 \rangle \rightarrow |F'=3 \rangle$  transition. A Gaussian beam with power of about 30 mW, red detuning of approximately 67 GHz from $|F=2 \rangle \rightarrow |F'=3 \rangle$  transition, and diameter of around 1.8 mm (forming potential well with depth about 15-30$T_{\rm r}$, where $T_{\rm r} \approx 0.362$ $\upmu$K is the recoil temperature), is incident into the vacuum chamber, reflected by a mirror to form a 1-dimension optical lattice at the center of the MOT. A Charge-Coupled Device (CCD) camera is mainly used to observe the atom cloud during operation and measure the atomic temperature. The light used for detection of time-of-flight (TOF) signal is shaped by a 100~mm$\times$1~mm aperture. When atom cloud passes through this region, the resonant fluorescence could be detected by a photonic detector (PD) and conveyed to the host computer by a 16-bit data acquisition card. 

\begin{figure}
	\centering
	\includegraphics[keepaspectratio=true]{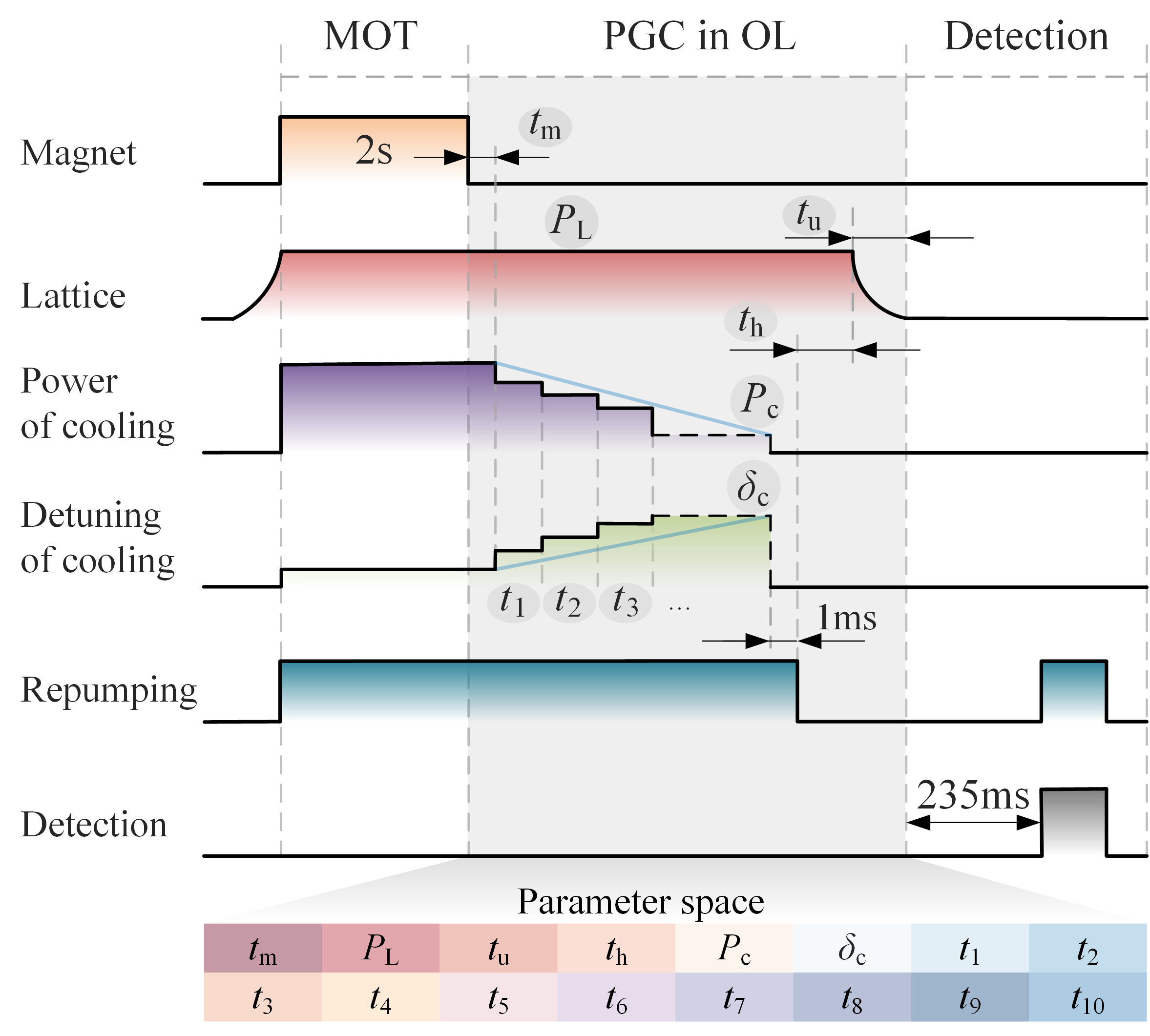}
	\caption{Timing sequence and parameters space. OL, optical lattice; $t_{\rm m}$, time to turn off magnetic coils; $P_{\rm L}$, power of lattice light; $t_{\rm u}$, duration of the optical lattice unloading; $t_{\rm h}$, holding time of optical lattice; $P_{\rm C}$, power of cooling light at the end of PGC; $\delta_{\rm C}$, detuning of cooling light at the end of PGC; $t_{i}$ ($i=1, 2, ..., 10$), the time step of PGC. The MOT loads for a duration of 2 seconds, followed by the starting of PGC after a delay of $t_{\rm m}$. The atoms are then pumped to the $|F=2 \rangle$ state by turning off the repumping light 1 ms later than the cooling light. After being released from lattice, the atoms undergo a 235 ms free-falling in vacuum chamber before reaching the detection zone.}
	\label{fig:timing}
\end{figure}
Setting reasonable parameters $\mathbf{X}$, one experiment cycle can be conducted following the timing sequence depicted in FIG.~\ref{fig:timing}. Consequently, the TOF signal can be obtained on the host computer. The atomic temperature $T_{\rm a}$ is acquired by fitting TOF signal in velocity domain (see Appendix~\ref{app:tof}), which is subsequently used to calculate the objective function $O(\mathbf{X})$. A pairwise observed data point consisting of scaled parameters $\mathbf{X}$ and corresponding $O(\mathbf{X})$ would be used to fit the surrogate model, which is Gaussian process in this study. MHCS-BO is utilized to predict the optimal parameters for next experiment cycle. The above process iterates until reaching the termination criterion. To finely control the process of PGC, we discarded the linear changing of cooling light parameters in traditional PGC (blue line). Instead, the PGC interaction time is divided into 10 steps, and each step $t_i (i = 1, 2, …, 10)$ is variable to be optimized. The initial power and detuning of cooling light in PGC are the same as that in MOT stage, and are subsequently changed to $P_{\rm c}$ and $\delta_{\rm c}$ at the termination of PGC stage. There is a time delay $t_{\rm m}$ between MOT and PGC. Through adjusting $t_i$, $P_{\rm c}$ and $\delta_{\rm c}$, the power and detuning can have different sequence, achieving finely controlling of PGC process. After the completion of PGC, only the lattice light continues to hold on for a certain time $t_{\rm h}$ and then unload adiabatically within time $t_{\rm u}$ to achieving lower $T_{\rm a}$. However, if $t_{\rm h}$ is too long, atoms will be accelerated due to gravity and the lattice, leading to a decrease in atom number $N_{\rm a}$ within potential well. Additionally, the power of lattice light $P_{\rm L}$ can affect the well depth and scattering rate. A deep well results in more confined atoms, but the velocity distribution of atoms will be widened, resulting in higher $T_{\rm a}$. Conversely, a shallow well corresponds to fewer $N_{\rm a}$ and lower $T_{\rm a}$. Moreover, a higher the scattering rate, which means a higher probability of interaction between atoms and photons, would elevate the probability of atom heating. Thus, $t_{\rm h}$ and $P_{\rm L}$ are key parameters that needs to be optimized and balanced.

In summary, the parameters that need optimization are $\mathbf{X}=\{P_{\rm c}, \delta_{\rm c}, t_i, t_{\rm m}, t_{\rm u}, t_{\rm h}, P_{\rm L}\}$ (16 dimensions). As $P_{\rm c}$ is controlled by an acousto-optic modulator (AOM), its value can be directly controlled by setting the attenuation ($Att$) of radio-frequency signal. The $P_{\rm L}$ is controlled by another AOM with its radio-frequency signal provided by an arbitrary waveform generator, and the signal amplitude is modulated by the voltage $V_{\rm L}$ provided by the controller. We use $Att$ and $V_{\rm L}$ to replace $P_{\rm c}$ and $P_{\rm L}$ (see \href{run:./supplemental_material.pdf}{supplemental material}). Therefore, the actual parameters for optimization are $\mathbf{X}_{\rm a}=\{Att, \delta_{\rm c}, t_i, t_{\rm m}, t_{\rm u}, t_{\rm h}, V_{\rm L}\}$. If all parameters are optimized together in a coupled manner, it could easily lead to poor signals, misleading the optimization direction. Hence, we first turn off the optical lattice and optimize the PGC-related parameters  $\mathbf{X}_{\rm P}=\{Att, \delta_{\rm c}, t_i, t_{\rm m}\}$to initially cool the atoms to below 5 $\upmu$K. Subsequently, we turn on the lattice and optimize the lattice-related parameters $\mathbf{X}_{\rm L}= \{t_{\rm h}, t_{\rm m}, t_{\rm u}, V_{\rm L}\}$ to achieve deep cooling of atoms. It is noteworthy that the magnetic field delay parameter $t_{\rm m}$ also has some impact on the cooling effect in optical lattice, thus it is included in $\mathbf{X}_{\rm L}$. The attenuation coefficient, light power, time, and voltage are measured in dB, mW, ms, and V, respectively.

\begin{figure}
	\centering
	\includegraphics[keepaspectratio=true]{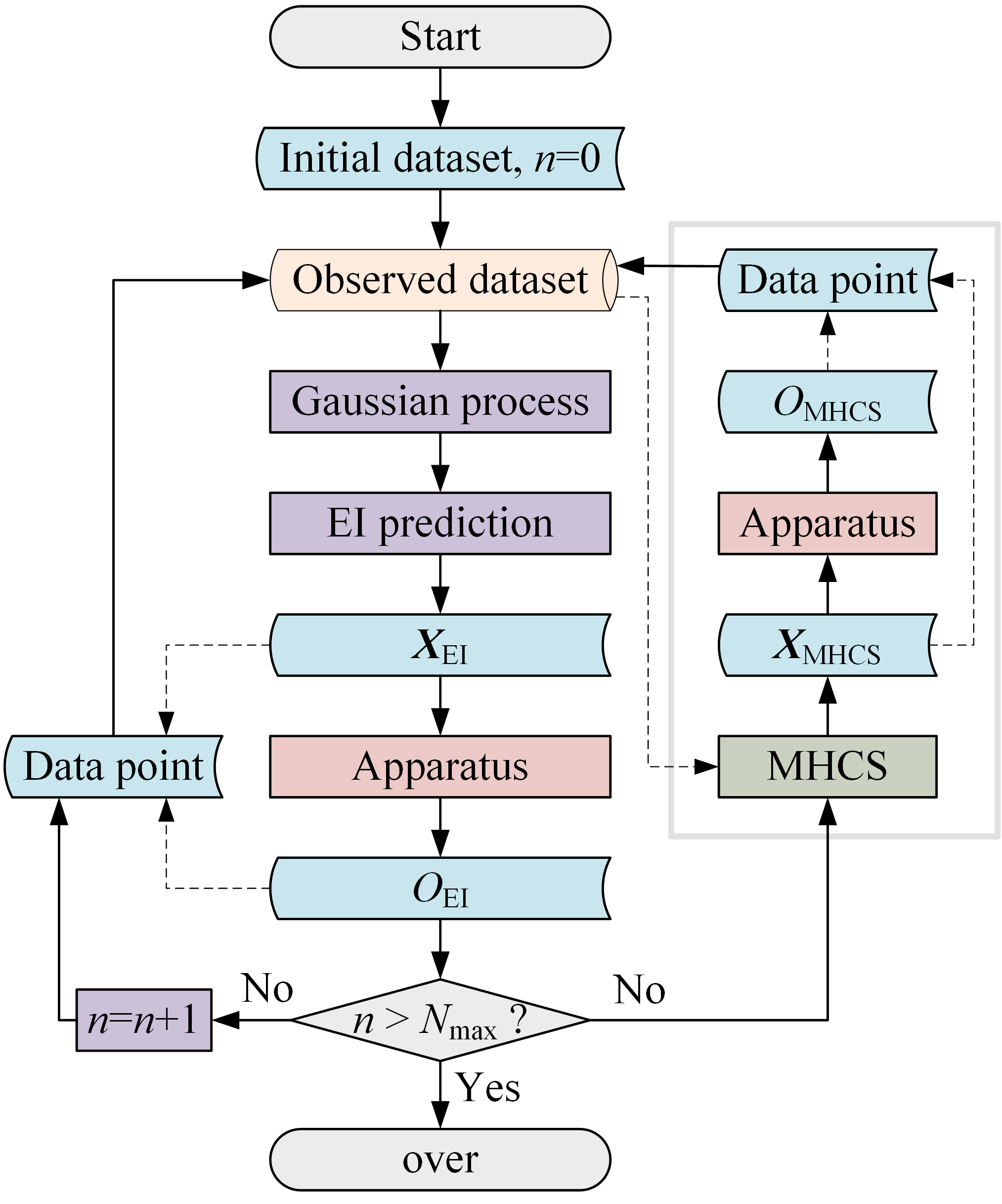}
	\caption{Flowchart of MHCS-BO. The light blue blocks represent the data generated during optimization process, the red blocks indicate implementing parameters to apparatus and conducting one experiment, the purple blocks depict critical steps in BO, and the gray frame indicates MHCS prediction.}
	\label{fig:alg}
\end{figure}
\section{MHCS-BO algorithm}
\begin{figure}
	\centering
	\includegraphics[width=\columnwidth]{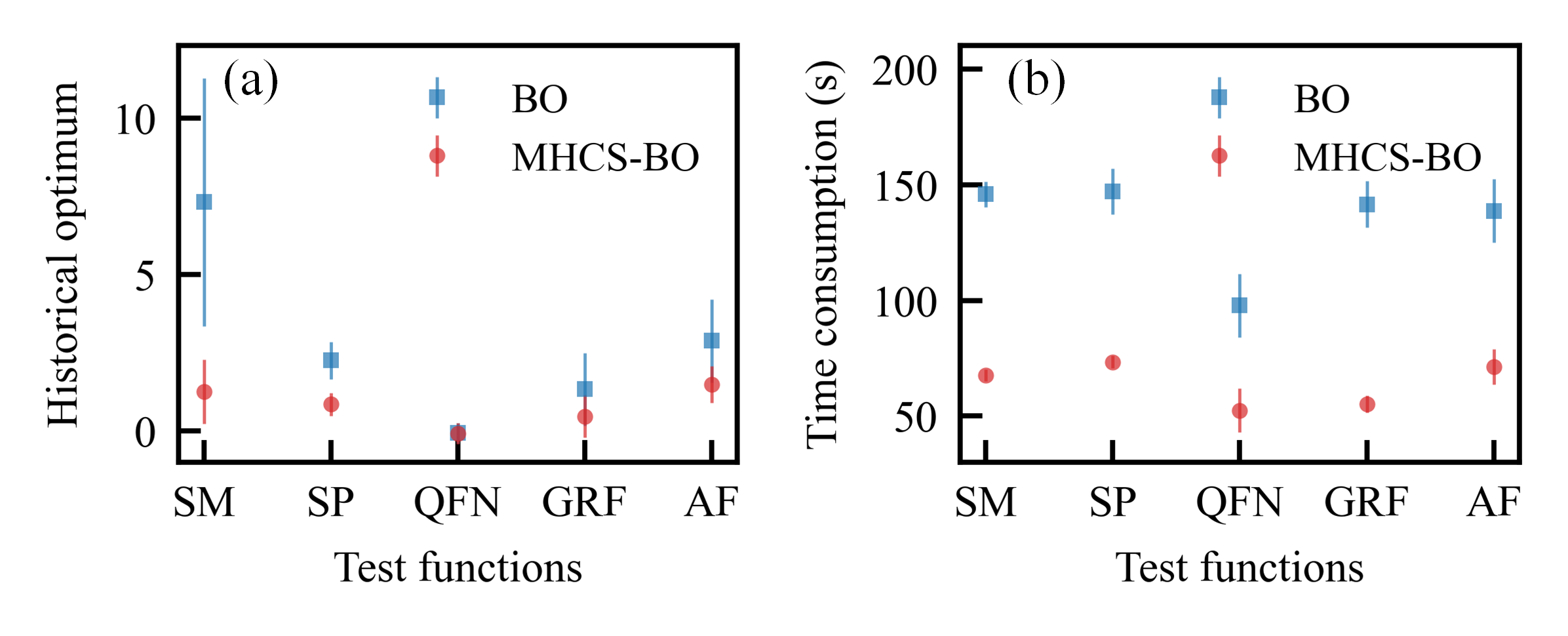}
	\caption{Performance comparison between MHCS-BO and BO under 5 typical test functions. The historical optimum (a) and time consumption (b) under different test functions are displayed. Each data is averaged by 9 repetitions, and the error bars indicate one standard deviation. SM, Sphere Model; SP, Schwefel’s Problem 2.21; QFN, Quartic Function i.e. Noise; GRF, Generalized Rastrigin’s Function; AF, Ackley’s Function.}
	\label{fig:algrithm_valid}
\end{figure}

MHCS-BO is proposed to find the optimal parameters as shown in FIG.~\ref{fig:alg}. The Gaussian process is fitted initially by the dataset which could be obtained by prior data or randomly sampled in parameter space. The prediction stage is mainly composed of two steps: (1) prediction based on EI function; (2) prediction based on MHCS. The corresponding optimal point $\mathbf{X}_{\rm EI}$ and $\mathbf{X}_{\rm MHCS}$ would be implemented in the apparatus, yielding the objective function $O_{\rm EI}$ and $O_{\rm MHCS}$ respectively, all of which would be added to the observed dataset. To balance $T_{\rm a}$ and $N_{\rm a}$, the objective function is defined as $O(\mathbf{X}) = T_{\rm a}/A_{\rm s}^b$, where $A_{\rm s}\propto N_{\rm a}$ represents the amplitude of TOF signal, and $b$ is a balancing coefficient within range of 0.1-1. A higher $b$ prioritizes maximizing atom numbers, whereas a lower $b$ emphasizes minimizing atom temperature. To ensure sufficient atom numbers while pursuing deep cooling, we empirically set $b$=0.5, which strikes a balanced compromise between these competing goals. As the iteration process running, the fitted Gaussian process would be more accurate to reflect the mapping relationship between the experimental parameters and the corresponding $O(\mathbf{X})$. To assess the performance and generalizability of MHCS-BO, we apply it to 5 standard test functions \cite{yao1999jul} ( see Appendix \ref{app:mhcs_test}). Here, the optimization accuracy is characterized by the historical optimum values recorded during iterative episodes. The results reveal that MHCS-BO outperforms traditional BO, with nearly double optimization speed and superior prediction accuracy as shown in FIG.~\ref{fig:algrithm_valid}.
\begin{figure*}
	\centering
	\includegraphics{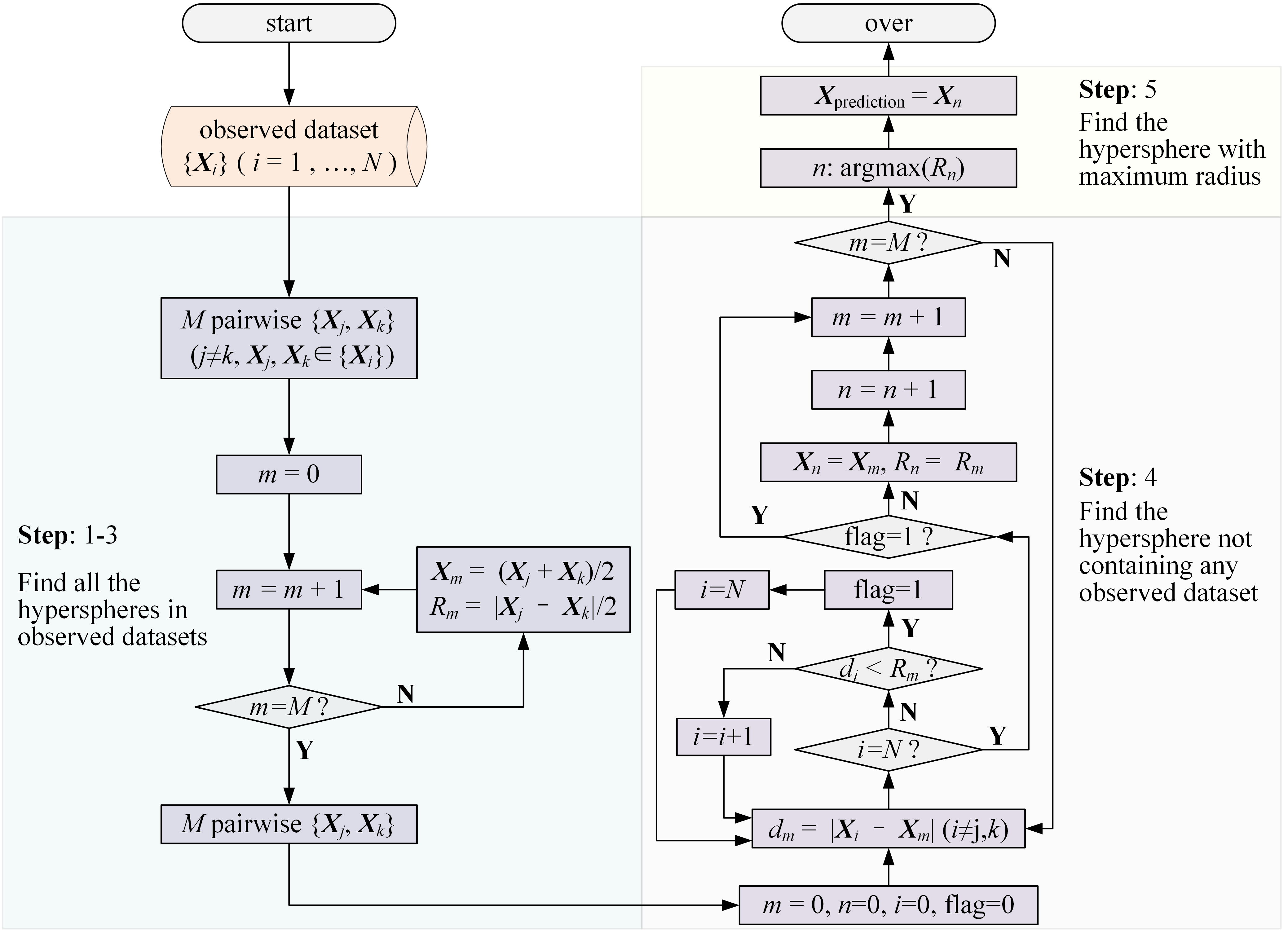}
	\caption{Flowchart of MHCS algorithm. Yellow block represents the observation dataset, light purple and gray blocks represent the parameters and key processes, respectively.}
	\label{fig:MHCS}
\end{figure*}

In traditional BO using EI prediction (see Appendix \ref{app:TBO}), the predicated optimal points may cluster in some certain regions, leaving large non-sampling regions. This situation is quite common in high-dimensional spaces, leading to incomplete observation. Consequently, the inaccurate mapping from surrogate model to apparatus reduces the algorithm’s performance. The core idea of MHCS is to identify the largest sparse region based on the current observed dataset and sample one point within that region. Here, we first give a brief introduction to the concept of hypersphere. Assuming that there are two points in the 3-dimensional space, $\mathbf{X}_1$ and $\mathbf{X}_2$, a sphere can be generated with the line connecting the two points as diameter. Similarly, in spaces with more than 3 dimensions as the problem in this study, the sphere becomes a hypersphere. Now we will give a more rigorous mathematical description of the hypersphere. The midpoint between $\mathbf{X}_1$ and $\mathbf{X}_2$ is defined as $\mathbf{X}_{\rm m} = (\mathbf{X}_1 + \mathbf{X}_2)/2$. The Euclidean distance between them is denoted as $D = | \mathbf{X}_1 - \mathbf{X}_2|$. The surface in the entire parameter space composed of points equidistant from the midpoint at a distance of $D/2$ is called a hypersphere with $\mathbf{X}_{\rm m}$ as the center and a diameter of $D$.

As depicted in FIG. \ref{fig:MHCS}, the main steps of MHCS algorithm are as follows: 

\textbf{Step1}: To ensure consistency across scales for subsequent processing, all observed data points are scaled within the range of 0 to 1.

\textbf{Step2}: For the current set of observed $N$ data points, $\mathbf{X}_i$ ($i=0, 1, …, N-1$), identify all possible pairwise combinations $\{\mathbf{X}_j$, $\mathbf{X}_k\}$ ($j \neq k; \mathbf{X}_j, \mathbf{X}_k\in \{\mathbf{X}_i\}$).

\textbf{Step3}: Calculate their respective centers $\{\mathbf{X}_{m}\}$ and radius $\{R_m\}$, where $m$=0, 1, …, $M$.

\textbf{Step4}: For each combination $\{\mathbf{X}_j, \mathbf{X}_k\}$ corresponding to a hypersphere, iterate through all observed parameters $\{\mathbf{X}_i\}$ $(i\neq j, k)$. If any observed parameter satisfies |$\mathbf{X}_m - \mathbf{X}_i$| > $R_c$, it indicates that no observed dataset is in this hypersphere, which is then selected as a candidate hypersphere.

\textbf{Step5}: Choose the candidate hypersphere with the largest radius , and the corresponding center $\mathbf{X}_m$ is the recommended point of the algorithm.

\section{Results}
\begin{figure*}
	\centering
	\includegraphics[keepaspectratio=true]{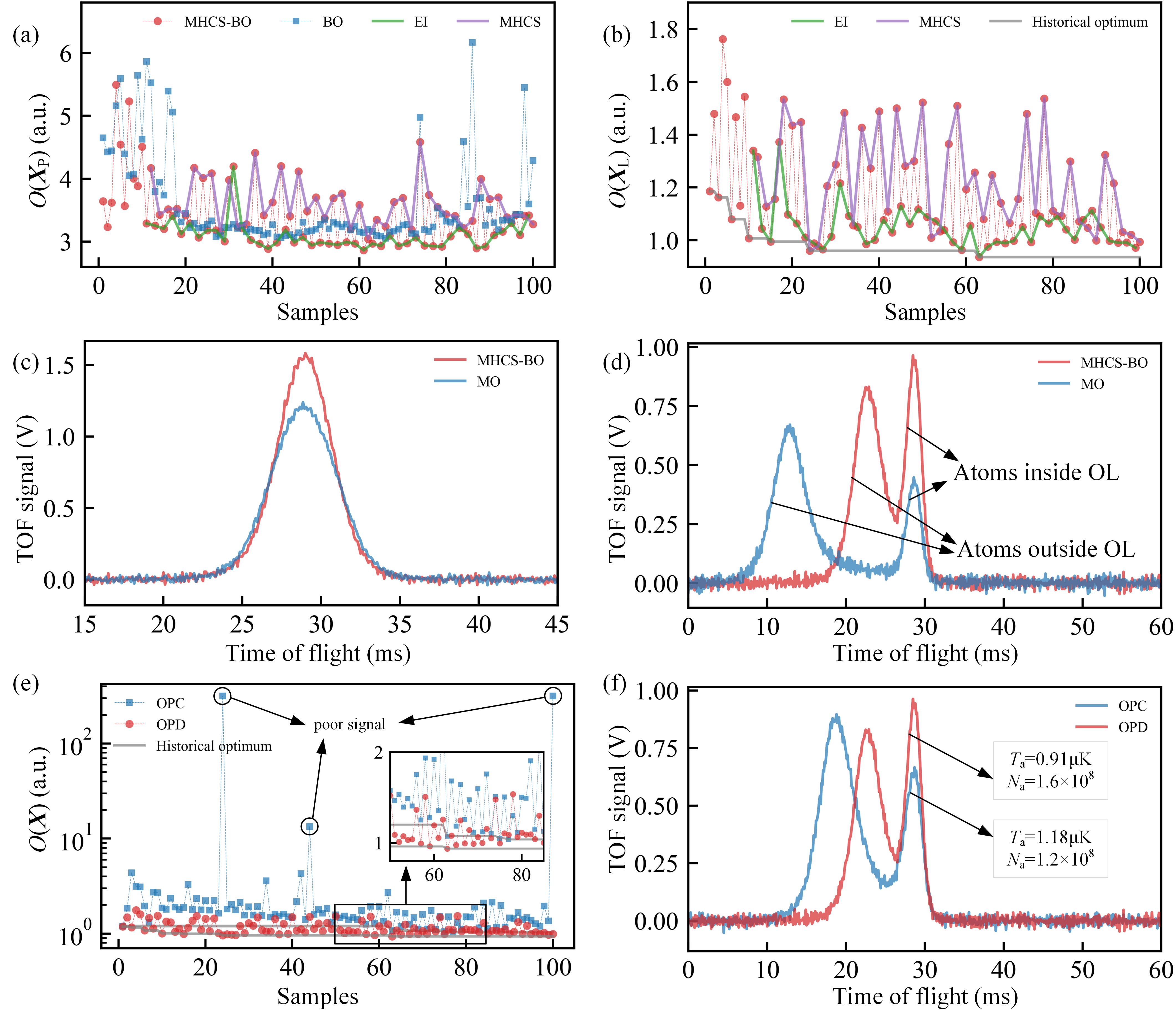}
	\caption{Optimization of PGC and optical lattice. (a) The optimization of PGC using two different BO algorithms. (b) The MHCS-BO algorithm is used to optimize the 4 parameters of optical lattice. (c) and (d) show the TOF signals of cold atom clouds under the optimal parameters when optical lattice is closed and open, with the red lines indicating the MHCS-BO results and the blue lines representing the MO results. (e) The BO process under OPC and OPD conditions, where the gray lines represent their respective historical optimum. (f) The best optimization results under OPC and OPD conditions.}
	\label{fig:opt}
\end{figure*}

Here, as for the 16 parameters in our experiment, we first provide an intuitive comparison of the optimization efficiency among three methods: MHCS-BO, manual optimization (MO), and brute-force search. MHCS-BO requires decoupling and optimizing the parameters step by step, running a total of 200 experiments, taking about 15 minutes. The MO similarly comprised two steps. For PGC, only the cooling laser power and detuning at the termination stage of PGC were adjusted following the light-blue sequence in FIG.~\ref{fig:timing}. Cooling performance is assessed via fluorescence images of the atom cloud after 10-25 ms of free fall, where smaller and brighter images corresponded to higher atom numbers and lower temperatures. Based on the optimized PGC configuration, the optical lattice parameters are then optimized by observing the TOF signal similar to that in FIG.~\ref{fig:opt}(d). The process was inherently dependent on the prior experience and less efficient than algorithmic approaches. Even MO is operated by an experienced operator familiar with the system, it typically takes 4-6 hours to obtain satisfactory results. In the context of grid brute-force search, employing a parameter step size of $P_{\rm range}/3$ ($P_{\rm range}$ is the parameter range), which denotes a coarse granularity, necessitates conducting an impractical number of $3^{16}$ experiments, equivalent to an estimated duration of approximately 6 years. As a result, MHCS-BO is appropriate to our experiment.

The optimization results under different conditions are illustrated in FIG.~\ref{fig:opt}. The impact of BO and MHCS-BO on the optimization process of PGC is compared in FIG.~\ref{fig:opt}(a). The iterations for BO and MHCS-BO are set to 100 and 50 respectively, yielding 100 sampling points. It can be observed that in the vast majority of iterations, the EI prediction line in MHCS-BO is significantly lower than that in BO, indicating that MHCS-BO can obtain promising parameters. This implies that exploration of non-sampling sparse regions by the MHCS effectively enhances the completeness of observed dataset, bringing a more precise mapping from the model to apparatus, result in greatly improving EI prediction performance. MHCS-BO and BO reach the historical global optimum at the 61st and 66th sample respectively.

The historical global optimal parameters for MHCS-BO are $\mathbf{X}^*_{\rm P}=\{$31.5, 188.21, 10.0, 6.42, 2.37, 5.25, 0.9, 0.42, 9.96, 7.76, 1.04, 0.3, 6.87$\}$. Setting the PGC parameters $\mathbf{X}_{\rm P} = \mathbf{X}^*_{\rm P}$, $\mathbf{X}_{\rm L}$ is optimized keeping the lattice light holding on. The results is depicted in FIG.~\ref{fig:opt}(b). The majority of the best values are generated by EI predictions which is similar to FIG.~\ref{fig:opt}(a). This suggests that MHCS predictions serve as effective supplements to EI predictions, which plays a role to explore the blind areas in the parameter space, thereby enhance the overall performance of the model. However, there is still one MHCS prediction point appearing in the historical best value (gray solid line), indicating that the MHCS algorithm has a certain probability of exploring potential optimal values that the EI algorithm cannot predict, highlighting the necessity of the MHCS algorithm. The entire iterative process reaches the historical global optimum at the 63nd sample with $\mathbf{X}^*_{\rm L}$ = $\{4.75, 9.76, 0.6, 4.16\}$. It is noteworthy that the magnetic field delay $t_{\rm m}$ here is different from that in $\mathbf{X}^*_{\rm P}$, changing from 6.87 ms to 4.75 ms. The last optimal parameters $\mathbf{X}^*=\mathbf{X}^*_{\rm P}(1:12)\cup \mathbf{X}^*_{\rm L}$.

\begin{figure}
	\includegraphics{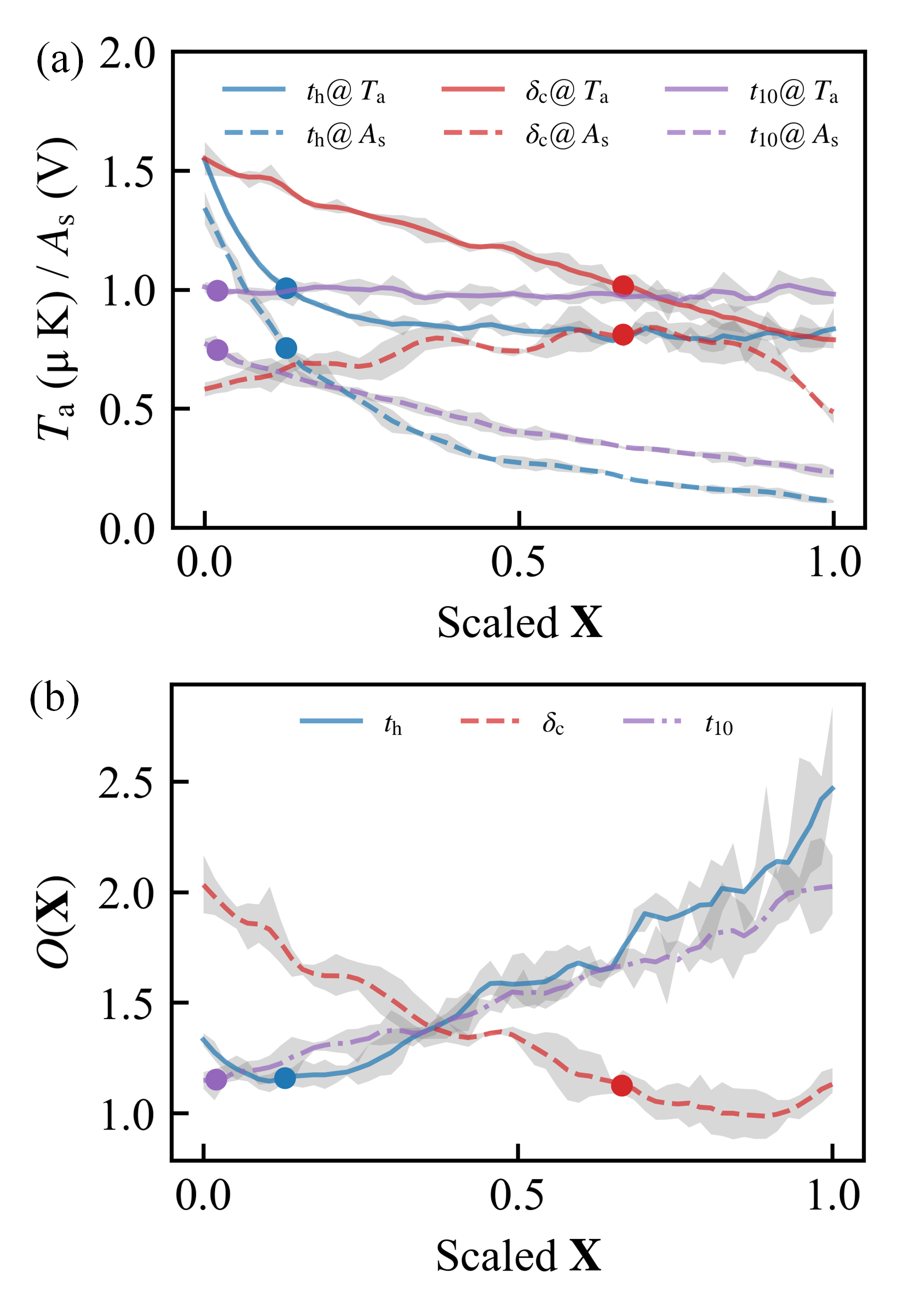}
	\caption{Key parameter scans under optimal settings. When one parameter changes, the remaining ones are kept at $\mathbf{X}^*$, indicated by solid dots in corresponding colors. Each data represents the average of 3 measurements, and the gray area donates one standard deviation. (a) The changing of $T_{\rm a}$ (solid line) and $A_{\rm s}$ (dashed line) versus scaled parameters. (b) Variation of $O(\mathbf{X})$ versus scaled $\mathbf{X}$.}
	\label{fig:dp}
\end{figure}
The optimal $T_{\rm a}$ and $N_{\rm a}$ of PGC obtained by MHCS-BO and MO are 3.1 $\upmu \rm K$, 4.3 $\upmu \rm K$, and $4.2\times10^8$, $3.6\times10^8$ as shown in FIG.~\ref{fig:opt}(c), respectively. Although MO can also achieve a relatively acceptable temperature and atom number, the optimization process relies on the intuition and familiarity of the system. In other words, different individuals may obtain different optimization results, leading to a degree of subjectivity. In contrast, MHCS-BO executes optimization more objectively and accurately based on $O(\mathbf{X})$, resulting in significantly higher optimization metrics compared to MO. With optical lattice holding on and the parameters setting to $\mathbf{X}^*$, the TOF signals is compared between MHCS-BO and MO. It can be seen in FIG.~\ref{fig:opt}(d) that due to the dipole trap of optical lattice overcoming some of the effects from gravity, the atoms inside optical lattice fall later than those outside. Consequently, the peak of TOF signal exhibits a relative lag. Moreover, the TOF signal for atoms confined within optical lattice displays significantly narrower width, indicating a better cooling effect. The optimal $T_{\rm a}$ and $N_{\rm a}$ inside optical lattice for MHCS-BO and MO are 0.91 $\upmu \rm K$, 0.82 $\upmu \rm K$, and $1.6\times10^8$ and $7.1\times10^7$, respectively. With the assistance of optical lattice, MO also achieves acceptable atomic temperature, but the atom number is lower than that of MHCS-BO. This is because we have set a comprehensive optimization metric that balances both $T_{\rm a}$ and $N_{\rm a}$, rather than solely optimizing the atomic temperature.

In order to verify the necessity of parameters decoupling, we compare the optimization under parameters coupling (OPC) and decoupling (OPD) as shown in FIG.~\ref{fig:opt}(e). Overall, $O(\mathbf{X})$ of OPC is generally worse than that of OPD. In our experiments, $T_{\rm a}$ cannot be calculated if TOF signal is poor (black circle). In such instance, it is directly assigned the value of 100. As a result, the mapping accuracy between the model and apparatus may be compromised, potentially misleading the optimization direction and slowing down the optimization efficiency. It can be seen that some poor TOF signals are generated in some samples under OPC. Conversely, such signal cannot be found in OPD. The optimal $T_{\rm a}$ and $N_{\rm a}$ inside optical lattice from OPC are 1.18 $\upmu \rm K$ and $1.2\times10^7$ respectively as depicted in FIG.~\ref{fig:opt}(f), even worse than the ones from MO. Additionally, $O(\mathbf{X})$ in OPC fluctuates more significantly than OPD, indicating more susceptible to the environment. Therefore, the overall performance of OPD surpasses that of OPC.

Considering the existence of systemic error and random error, the actual experimental parameters may deviate from the setting value. Therefore, it is crucial to ensure that our experimental results, generated by optimal parameters with a few deviations, are proximal to the optimum. Each parameter is scanned under the optimum ($\mathbf{X}^*$). Here, only the three parameters that have the greatest impact on cooling effects are analyzed as shown in FIG.~\ref{fig:dp}. The results for other parameters can be seen in \href{run:./supplemental_material.pdf}{supplemental material} for details. As the holding time $t_{\rm h}$ increases, both $T_{\rm a}$ and $A_{\rm s}$ decrease (blue line). This phenomenon aligns with physical intuition, because the dipole force, scattering force, and gravity all act on the atoms within optical lattice. As a result, some atoms escape from the potential well, leading to a decrease in the number of atoms inside the trap. The remaining atoms are those with kinetic energy less than well depth, which meaning a lower atomic temperature. With the further extension of  $t_{\rm h}$, $T_{\rm a}$ will gradually stabilize. When $\delta_{\rm c}$ gradually increases (red line), there is no apparent pattern in the changes of $T_{\rm a}$ and $A_{\rm s}$. Overall, when $\delta_{c}$ is not too large, $T_{\rm a}$ decreases with increasing $\delta_{c}$, while $A_{\rm s}$ exhibits the opposite trend. $t_{10}$ represents the final time step of PGC. As this value increases, $A_{\rm s}$ decreases gradually, while the change in $T_{\rm a}$ is less pronounced (purple line). This phenomenon indicates that the atoms have already been sufficiently cooled in the preceding time steps. Therefore, extending the value of $t_{10}$ is equivalent to increasing $t_{\rm h}$, resulting in a corresponding alignment between the trends in $T_{\rm a}$ and $A_{\rm s}$ with the analysis of $t_{\rm h}$. Since $O(\mathbf{X})$ is based on $T_{\rm a}$ and $A_{\rm s}$, the lowest atomic temperature was not set as the optimization target. Instead, a balance between them was considered. The position of optimal parameter can vividly reflect this, indicating that the objective function has achieved the intended purpose. As shown in FIG.~\ref{fig:dp}(b), the changes of $O(\mathbf{X})$ in the vicinity of the optimal point are relatively smooth, without abrupt gradient shifts. This implies that the optimal parameters $\mathbf{X}^*$ predicted by MHCS-BO do not coincidentally fall on a singular optimal point and demonstrate a certain degree of robustness, highlighting the superiority of the algorithm proposed in this paper for optimizing multiple parameters in PGC.
\begin{figure}
	\centering
	\includegraphics[keepaspectratio=true]{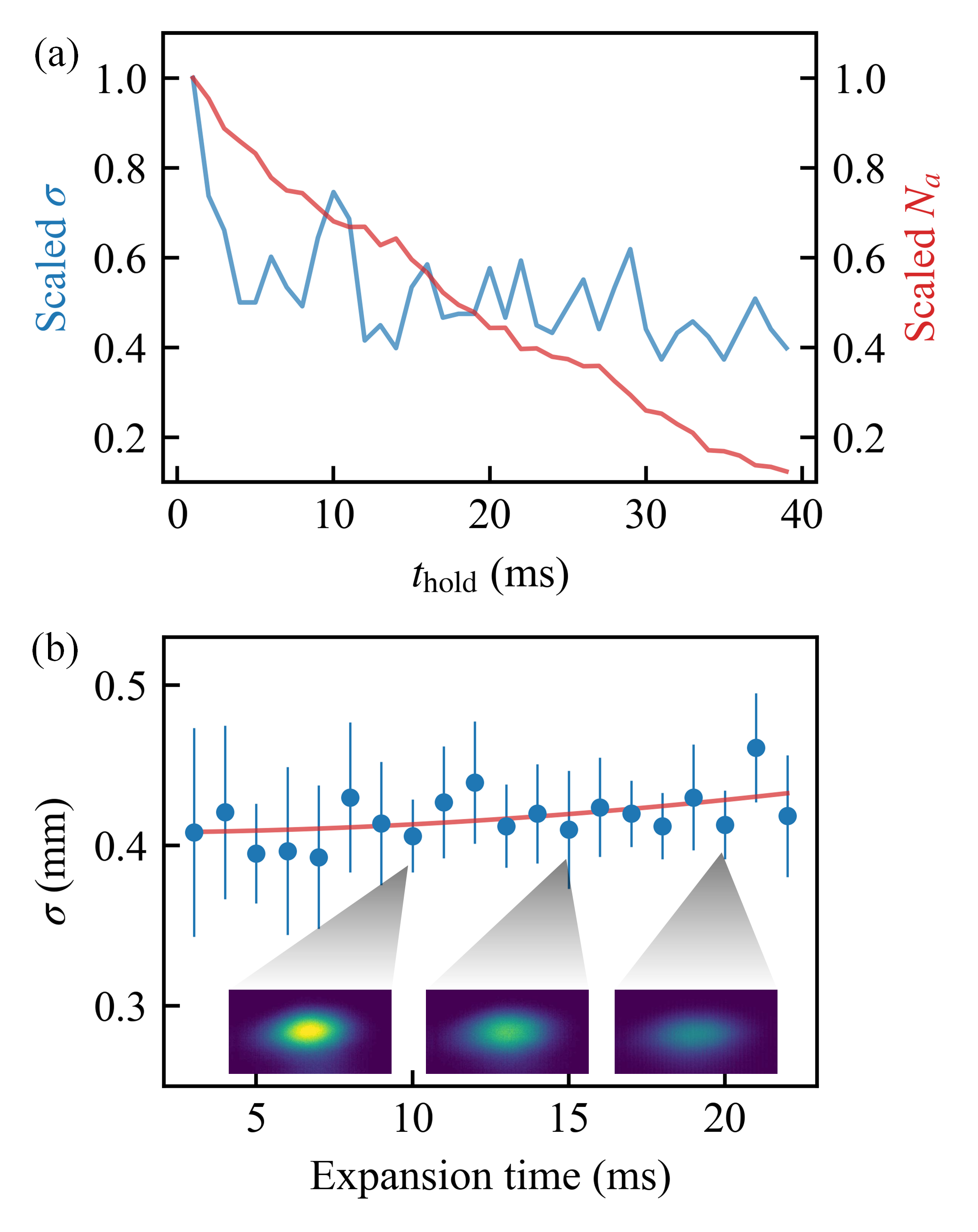}
	\caption{(a) scaled $\sigma$ and $N_{\rm a}$ versus $t_{\rm h}$. $\sigma$ is the Gaussian radii of the atom cloud, which is obtained by fitting the fluorescence image along vertical direction. A smaller $\sigma$ means a corresponding lower atomic temperature. $N_{\rm a}$ is calculated by integrating the counts of fluorescence image. (b) Mean radii of atom cloud versus expansion time. Each data is averaged by 10 repetitions, and the error bars indicate the standard deviation. The insets is fluorescence images of atom cloud taken during ballistic expansion. The loading time is adjusted to 3 seconds in order to enhance the detection signal, while the exposure time of CCD camera is set at 0.5 ms.}
	\label{fig:Ty}
\end{figure}

In order to further explore lower $T_{\rm a}$, we increase $t_{\rm h}$ under the optimal parameters obtained from MHCS-BO. However, as $t_{\rm h}$ increased, the number of atoms gradually decreased.This would lead to some problems in the measurement of atom temperature. On one hand, the weakened atom fluorescence degrades TOF signal fitting precision. On the other hand, the reduced atom cloud size compromises temperature measurement accuracy, because the thickness of the probe light could not be omitted (see Appendix D). Therefore, we increase the MOT loading time to 3 seconds to obtain more atoms, and employ the ballistic expansion method to precisely measure atomic temperature. By setting $t_{\rm h}$ to 20 ms, a lower $T_{\rm a }$ could be achieved without a significant decrease in $N_{\rm a}$ as depicted in FIG.~\ref{fig:Ty}(a). Under this condition, images of atom cloud at different expansion time $t$ were captured by a CCD camera. The relationship between radius of atom cloud $\sigma$ and expansion time $t$ could be expressed as $\sigma^2=\sigma_0^2 + (k_{\rm B}T_{\rm a}/m)t^2$ \cite{yuan2024jan}. Where $\sigma_0$ is the initial radius of the atom cloud, $k_{\rm B}$ is the Boltzmann constant, and $m$ is the mass of $^{87}$Rb atom. By fitting the above equation as shown in  FIG.~\ref{fig:Ty}(b), $T_{\rm a}$ can be obtained as 0.4$\pm$0.2 $\upmu$K, very close to the recoil temperature.

\begin{figure}[ht!]
	\centering
	\includegraphics[keepaspectratio=true]{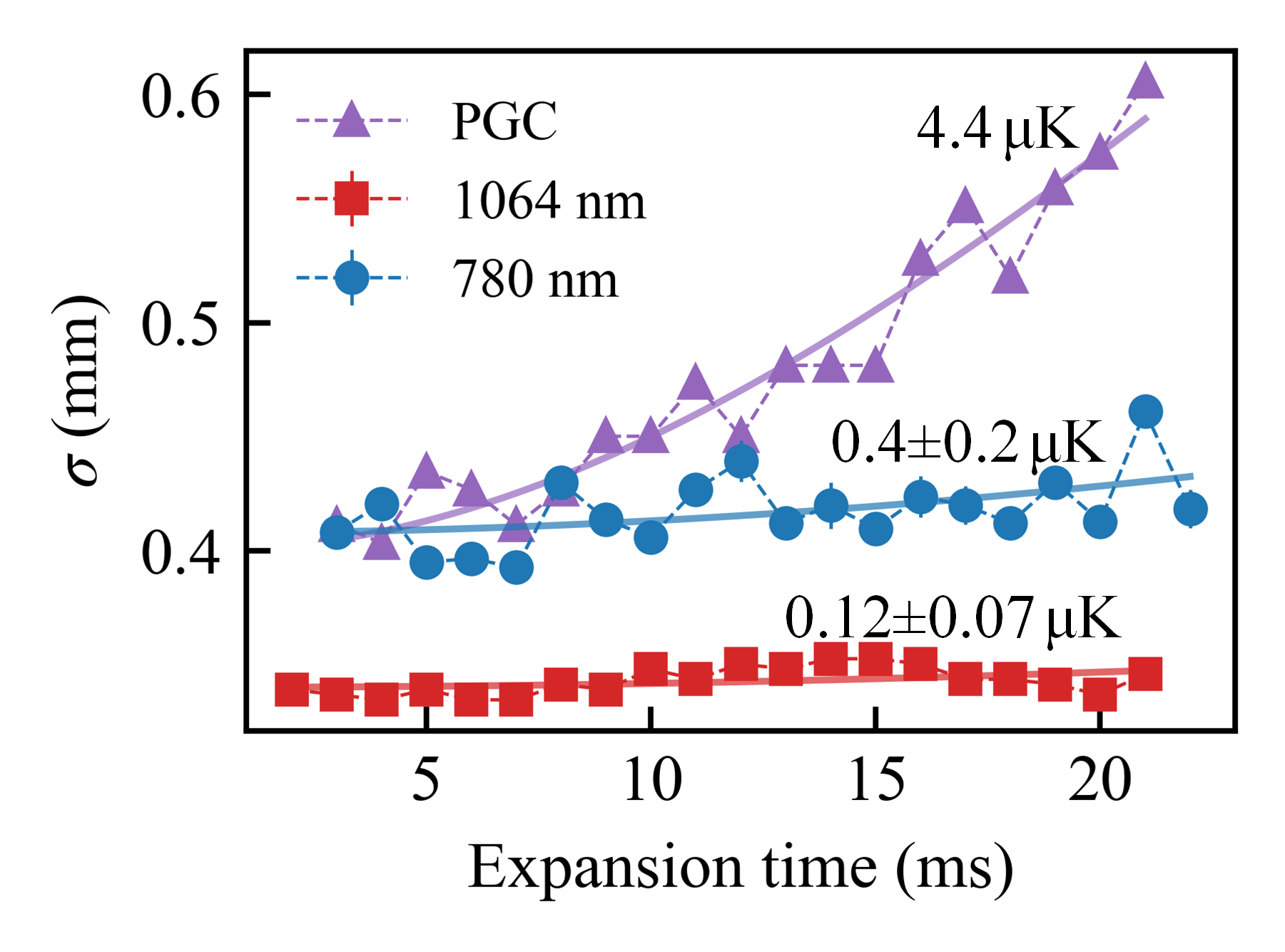}
	\caption{Atom temperature measurements under optical lattices with different wavelengths compared to PGC. The atom cloud under PGC exhibits significant spatial expansion, and no repeated measurements were conducted. The data under 1064 nm optical lattice represent averages of 5 independent measurements, with error bars indicating one standard deviation.}
	\label{fig:lattice1064}
\end{figure}
To further reduce scattering rates and extend atom lifetime, we also attempt to use a 1064 nm laser as the lattice light source. The vertical temperature obtained under three experimental conditions are shown in FIG.~\ref{fig:lattice1064}. Following the same MHCS-BO optimization protocol for 1064 nm optical lattice, we first identified an optimal parameter set $\mathbf{X}^*$ yielding a vertical temperature of approximate 0.6 $\upmu$K. Subsequent extension of the lattice holding time to 200 ms enabled the preparation of a cold atom ensemble containing $2\times10^8$ atoms with vertical temperature $0.12 \pm 0.07$ $\upmu$K. This temperature lies approximately three times below the recoil limitation, which holds significant potential to enhance fringe contrast and atom in atom interferometer (see Appendix \ref{app:Interference}).

\section{conclusions}
In summary, we have provided the first demonstration of MHCS-BO used in preparing sub-$\upmu$K cold atoms through PGC assisted with optical lattice. MHCS-BO inherits merit of BO, enhances the prediction accuracy and accelerates experimental parameters optimization. Compared to BO, MO, and grid brute-force search, MHCS-BO exhibits significant superiority in fast predicting optimal experimental parameters. Within 15 minutes of optimization, the temperature of atoms in the 780 nm optical lattice is 0.4$\pm$0.2 $\upmu$K with the atom number of $10^8$. For the first time, hundreds of millions of cold atoms close to recoil temperature are created, and it can be extended to 3-dimension cooling by adding another lattice beam. One limitation of MHCS-BO however is that it cannot find the complete Pareto frontier in muti-objective optimization problems. Thus, future iteration could use multi-objective BO to overcome this limitation. Overall, this study demonstrates a new scheme to generate high quality sub-$\upmu$K cold atoms, providing insights for initial cooling of cold atom experiments, e.g., BEC, quantum computing and quantum simulation.
\begin{acknowledgments}
	This work was supported by National Natural Science Foundation of China (Grant No. 12404556, 62103426) and Hunan provincial major sci-tech program (2023zk1010).
\end{acknowledgments}

\bibliography{mybib}

\begin{thebibliography}{38}%
\makeatletter
\providecommand \@ifxundefined [1]{%
 \@ifx{#1\undefined}
}%
\providecommand \@ifnum [1]{%
 \ifnum #1\expandafter \@firstoftwo
 \else \expandafter \@secondoftwo
 \fi
}%
\providecommand \@ifx [1]{%
 \ifx #1\expandafter \@firstoftwo
 \else \expandafter \@secondoftwo
 \fi
}%
\providecommand \natexlab [1]{#1}%
\providecommand \enquote  [1]{``#1''}%
\providecommand \bibnamefont  [1]{#1}%
\providecommand \bibfnamefont [1]{#1}%
\providecommand \citenamefont [1]{#1}%
\providecommand \href@noop [0]{\@secondoftwo}%
\providecommand \href [0]{\begingroup \@sanitize@url \@href}%
\providecommand \@href[1]{\@@startlink{#1}\@@href}%
\providecommand \@@href[1]{\endgroup#1\@@endlink}%
\providecommand \@sanitize@url [0]{\catcode `\\12\catcode `\$12\catcode
  `\&12\catcode `\#12\catcode `\^12\catcode `\_12\catcode `\%12\relax}%
\providecommand \@@startlink[1]{}%
\providecommand \@@endlink[0]{}%
\providecommand \url  [0]{\begingroup\@sanitize@url \@url }%
\providecommand \@url [1]{\endgroup\@href {#1}{\urlprefix }}%
\providecommand \urlprefix  [0]{URL }%
\providecommand \Eprint [0]{\href }%
\providecommand \doibase [0]{https://doi.org/}%
\providecommand \selectlanguage [0]{\@gobble}%
\providecommand \bibinfo  [0]{\@secondoftwo}%
\providecommand \bibfield  [0]{\@secondoftwo}%
\providecommand \translation [1]{[#1]}%
\providecommand \BibitemOpen [0]{}%
\providecommand \bibitemStop [0]{}%
\providecommand \bibitemNoStop [0]{.\EOS\space}%
\providecommand \EOS [0]{\spacefactor3000\relax}%
\providecommand \BibitemShut  [1]{\csname bibitem#1\endcsname}%
\let\auto@bib@innerbib\@empty
\bibitem [{\citenamefont {Peters}\ \emph {et~al.}(1999)\citenamefont {Peters},
  \citenamefont {Chung},\ and\ \citenamefont {Chu}}]{peters1999}%
  \BibitemOpen
  \bibfield  {author} {\bibinfo {author} {\bibfnamefont {A.}~\bibnamefont
  {Peters}}, \bibinfo {author} {\bibfnamefont {K.~Y.}\ \bibnamefont {Chung}},\
  and\ \bibinfo {author} {\bibfnamefont {S.}~\bibnamefont {Chu}},\ }\bibfield
  {title} {\bibinfo {title} {Measurement of gravitational acceleration by
  dropping atoms},\ }\href {https://doi.org/10.1038/23655} {\bibfield
  {journal} {\bibinfo  {journal} {Nature}\ }\textbf {\bibinfo {volume} {400}},\
  \bibinfo {pages} {849} (\bibinfo {year} {1999})}\BibitemShut {NoStop}%
\bibitem [{\citenamefont {Jones}\ \emph {et~al.}(2006)\citenamefont {Jones},
  \citenamefont {Tiesinga}, \citenamefont {Lett},\ and\ \citenamefont
  {Julienne}}]{jones2006may}%
  \BibitemOpen
  \bibfield  {author} {\bibinfo {author} {\bibfnamefont {K.~M.}\ \bibnamefont
  {Jones}}, \bibinfo {author} {\bibfnamefont {E.}~\bibnamefont {Tiesinga}},
  \bibinfo {author} {\bibfnamefont {P.~D.}\ \bibnamefont {Lett}},\ and\
  \bibinfo {author} {\bibfnamefont {P.~S.}\ \bibnamefont {Julienne}},\
  }\bibfield  {title} {\bibinfo {title} {Ultracold photoassociation
  spectroscopy: {{Long-range}} molecules and atomic scattering},\ }\href
  {https://doi.org/10.1103/RevModPhys.78.483} {\bibfield  {journal} {\bibinfo
  {journal} {Rev. Mod. Phys.}\ }\textbf {\bibinfo {volume} {78}},\ \bibinfo
  {pages} {483} (\bibinfo {year} {2006})}\BibitemShut {NoStop}%
\bibitem [{\citenamefont {Safronova}\ \emph {et~al.}(2018)\citenamefont
  {Safronova}, \citenamefont {Budker}, \citenamefont {DeMille}, \citenamefont
  {Kimball}, \citenamefont {Derevianko},\ and\ \citenamefont
  {Clark}}]{safronova2018jun}%
  \BibitemOpen
  \bibfield  {author} {\bibinfo {author} {\bibfnamefont {M.~S.}\ \bibnamefont
  {Safronova}}, \bibinfo {author} {\bibfnamefont {D.}~\bibnamefont {Budker}},
  \bibinfo {author} {\bibfnamefont {D.}~\bibnamefont {DeMille}}, \bibinfo
  {author} {\bibfnamefont {D.~F.~J.}\ \bibnamefont {Kimball}}, \bibinfo
  {author} {\bibfnamefont {A.}~\bibnamefont {Derevianko}},\ and\ \bibinfo
  {author} {\bibfnamefont {C.~W.}\ \bibnamefont {Clark}},\ }\bibfield  {title}
  {\bibinfo {title} {Search for new physics with atoms and molecules},\ }\href
  {https://doi.org/10.1103/RevModPhys.90.025008} {\bibfield  {journal}
  {\bibinfo  {journal} {Rev. Mod. Phys.}\ }\textbf {\bibinfo {volume} {90}},\
  \bibinfo {pages} {025008} (\bibinfo {year} {2018})}\BibitemShut {NoStop}%
\bibitem [{\citenamefont {Cimini}\ \emph {et~al.}(2023)\citenamefont {Cimini},
  \citenamefont {Valeri}, \citenamefont {Polino}, \citenamefont {Piacentini},
  \citenamefont {Ceccarelli}, \citenamefont {Corrielli}, \citenamefont
  {Spagnolo}, \citenamefont {Osellame},\ and\ \citenamefont
  {Sciarrino}}]{cimini2023feb}%
  \BibitemOpen
  \bibfield  {author} {\bibinfo {author} {\bibfnamefont {V.}~\bibnamefont
  {Cimini}}, \bibinfo {author} {\bibfnamefont {M.}~\bibnamefont {Valeri}},
  \bibinfo {author} {\bibfnamefont {E.}~\bibnamefont {Polino}}, \bibinfo
  {author} {\bibfnamefont {S.}~\bibnamefont {Piacentini}}, \bibinfo {author}
  {\bibfnamefont {F.}~\bibnamefont {Ceccarelli}}, \bibinfo {author}
  {\bibfnamefont {G.}~\bibnamefont {Corrielli}}, \bibinfo {author}
  {\bibfnamefont {N.}~\bibnamefont {Spagnolo}}, \bibinfo {author}
  {\bibfnamefont {R.}~\bibnamefont {Osellame}},\ and\ \bibinfo {author}
  {\bibfnamefont {F.}~\bibnamefont {Sciarrino}},\ }\bibfield  {title} {\bibinfo
  {title} {Deep reinforcement learning for quantum multiparameter estimation},\
  }\href {https://doi.org/10.1117/1.AP.5.1.016005} {\bibfield  {journal}
  {\bibinfo  {journal} {Adv. Photonics}\ }\textbf {\bibinfo {volume} {5}},\
  \bibinfo {pages} {016005} (\bibinfo {year} {2023})}\BibitemShut {NoStop}%
\bibitem [{\citenamefont {Ciavarella}\ \emph {et~al.}(2023)\citenamefont
  {Ciavarella}, \citenamefont {Caspar}, \citenamefont {Singh},\ and\
  \citenamefont {Savage}}]{ciavarella2023}%
  \BibitemOpen
  \bibfield  {author} {\bibinfo {author} {\bibfnamefont {A.~N.}\ \bibnamefont
  {Ciavarella}}, \bibinfo {author} {\bibfnamefont {S.}~\bibnamefont {Caspar}},
  \bibinfo {author} {\bibfnamefont {H.}~\bibnamefont {Singh}},\ and\ \bibinfo
  {author} {\bibfnamefont {M.~J.}\ \bibnamefont {Savage}},\ }\bibfield  {title}
  {\bibinfo {title} {Preparation for quantum simulation of the (1 +
  1)-dimensional o(3) nonlinear $\sigma$ model using cold atoms},\ }\href
  {https://doi.org/10.1103/PhysRevA.107.042404} {\bibfield  {journal} {\bibinfo
   {journal} {Phys. Rev. A}\ }\textbf {\bibinfo {volume} {107}},\ \bibinfo
  {pages} {042404} (\bibinfo {year} {2023})}\BibitemShut {NoStop}%
\bibitem [{\citenamefont {Ott}\ \emph {et~al.}(2021)\citenamefont {Ott},
  \citenamefont {Zache}, \citenamefont {Jendrzejewski},\ and\ \citenamefont
  {Berges}}]{ott2021}%
  \BibitemOpen
  \bibfield  {author} {\bibinfo {author} {\bibfnamefont {R.}~\bibnamefont
  {Ott}}, \bibinfo {author} {\bibfnamefont {T.~V.}\ \bibnamefont {Zache}},
  \bibinfo {author} {\bibfnamefont {F.}~\bibnamefont {Jendrzejewski}},\ and\
  \bibinfo {author} {\bibfnamefont {J.}~\bibnamefont {Berges}},\ }\bibfield
  {title} {\bibinfo {title} {Scalable {{Cold-Atom Quantum Simulator}} for
  {{Two-Dimensional QED}}},\ }\href
  {https://doi.org/10.1103/PhysRevLett.127.130504} {\bibfield  {journal}
  {\bibinfo  {journal} {Phys. Rev. Lett.}\ }\textbf {\bibinfo {volume} {127}},\
  \bibinfo {pages} {130504} (\bibinfo {year} {2021})}\BibitemShut {NoStop}%
\bibitem [{\citenamefont {Halimeh}\ \emph {et~al.}(2022)\citenamefont
  {Halimeh}, \citenamefont {McCulloch}, \citenamefont {Yang},\ and\
  \citenamefont {Hauke}}]{halimeh2022}%
  \BibitemOpen
  \bibfield  {author} {\bibinfo {author} {\bibfnamefont {J.~C.}\ \bibnamefont
  {Halimeh}}, \bibinfo {author} {\bibfnamefont {I.~P.}\ \bibnamefont
  {McCulloch}}, \bibinfo {author} {\bibfnamefont {B.}~\bibnamefont {Yang}},\
  and\ \bibinfo {author} {\bibfnamefont {P.}~\bibnamefont {Hauke}},\ }\bibfield
   {title} {\bibinfo {title} {Tuning the topological $\theta$-angle in
  cold-atom quantum simulators of gauge theories},\ }\href
  {https://doi.org/10.1103/PRXQuantum.3.040316} {\bibfield  {journal} {\bibinfo
   {journal} {PRX Quantum}\ }\textbf {\bibinfo {volume} {3}},\ \bibinfo {pages}
  {040316} (\bibinfo {year} {2022})}\BibitemShut {NoStop}%
\bibitem [{\citenamefont {Cao}\ \emph {et~al.}(2024)\citenamefont {Cao},
  \citenamefont {Eckner}, \citenamefont {Lukin~Yelin}, \citenamefont {Young},
  \citenamefont {Jandura}, \citenamefont {Yan}, \citenamefont {Kim},
  \citenamefont {Pupillo}, \citenamefont {Ye}, \citenamefont {Darkwah~Oppong},\
  and\ \citenamefont {Kaufman}}]{cao2024oct}%
  \BibitemOpen
  \bibfield  {author} {\bibinfo {author} {\bibfnamefont {A.}~\bibnamefont
  {Cao}}, \bibinfo {author} {\bibfnamefont {W.~J.}\ \bibnamefont {Eckner}},
  \bibinfo {author} {\bibfnamefont {T.}~\bibnamefont {Lukin~Yelin}}, \bibinfo
  {author} {\bibfnamefont {A.~W.}\ \bibnamefont {Young}}, \bibinfo {author}
  {\bibfnamefont {S.}~\bibnamefont {Jandura}}, \bibinfo {author} {\bibfnamefont
  {L.}~\bibnamefont {Yan}}, \bibinfo {author} {\bibfnamefont {K.}~\bibnamefont
  {Kim}}, \bibinfo {author} {\bibfnamefont {G.}~\bibnamefont {Pupillo}},
  \bibinfo {author} {\bibfnamefont {J.}~\bibnamefont {Ye}}, \bibinfo {author}
  {\bibfnamefont {N.}~\bibnamefont {Darkwah~Oppong}},\ and\ \bibinfo {author}
  {\bibfnamefont {A.~M.}\ \bibnamefont {Kaufman}},\ }\bibfield  {title}
  {\bibinfo {title} {Multi-qubit gates and {{Schr{\"o}dinger}} cat states in an
  optical clock},\ }\href {https://doi.org/10.1038/s41586-024-07913-z}
  {\bibfield  {journal} {\bibinfo  {journal} {Nature}\ }\textbf {\bibinfo
  {volume} {634}},\ \bibinfo {pages} {315} (\bibinfo {year}
  {2024})}\BibitemShut {NoStop}%
\bibitem [{\citenamefont {Mol}\ \emph {et~al.}(2023)\citenamefont {Mol},
  \citenamefont {Esguerra}, \citenamefont {Meister}, \citenamefont {Bruschi},
  \citenamefont {Schell}, \citenamefont {Wolters},\ and\ \citenamefont
  {W{\"o}rner}}]{mol2023}%
  \BibitemOpen
  \bibfield  {author} {\bibinfo {author} {\bibfnamefont {J.-M.}\ \bibnamefont
  {Mol}}, \bibinfo {author} {\bibfnamefont {L.}~\bibnamefont {Esguerra}},
  \bibinfo {author} {\bibfnamefont {M.}~\bibnamefont {Meister}}, \bibinfo
  {author} {\bibfnamefont {D.~E.}\ \bibnamefont {Bruschi}}, \bibinfo {author}
  {\bibfnamefont {A.~W.}\ \bibnamefont {Schell}}, \bibinfo {author}
  {\bibfnamefont {J.}~\bibnamefont {Wolters}},\ and\ \bibinfo {author}
  {\bibfnamefont {L.}~\bibnamefont {W{\"o}rner}},\ }\bibfield  {title}
  {\bibinfo {title} {Quantum memories for fundamental science in space},\
  }\href {https://doi.org/10.1088/2058-9565/acb2f1} {\bibfield  {journal}
  {\bibinfo  {journal} {Quantum Sci. Technol.}\ }\textbf {\bibinfo {volume}
  {8}},\ \bibinfo {pages} {024006} (\bibinfo {year} {2023})}\BibitemShut
  {NoStop}%
\bibitem [{\citenamefont {Vovcenko}\ \emph {et~al.}(2021)\citenamefont
  {Vovcenko}, \citenamefont {Shishkov},\ and\ \citenamefont
  {Andrianov}}]{vovcenko2021}%
  \BibitemOpen
  \bibfield  {author} {\bibinfo {author} {\bibfnamefont {I.~V.}\ \bibnamefont
  {Vovcenko}}, \bibinfo {author} {\bibfnamefont {V.~{\relax Yu}.}\ \bibnamefont
  {Shishkov}},\ and\ \bibinfo {author} {\bibfnamefont {E.~S.}\ \bibnamefont
  {Andrianov}},\ }\bibfield  {title} {\bibinfo {title} {Dephasing-assisted
  entanglement in a system of strongly coupled qubits},\ }\href
  {https://doi.org/10.1364/OE.417294} {\bibfield  {journal} {\bibinfo
  {journal} {Opt. Express}\ }\textbf {\bibinfo {volume} {29}},\ \bibinfo
  {pages} {9685} (\bibinfo {year} {2021})}\BibitemShut {NoStop}%
\bibitem [{\citenamefont {Kasevich}\ and\ \citenamefont
  {Chu}(1991)}]{kasevich1991}%
  \BibitemOpen
  \bibfield  {author} {\bibinfo {author} {\bibfnamefont {M.}~\bibnamefont
  {Kasevich}}\ and\ \bibinfo {author} {\bibfnamefont {S.}~\bibnamefont {Chu}},\
  }\bibfield  {title} {\bibinfo {title} {Atomic interferometry using stimulated
  {{Raman}} transitions},\ }\href {https://doi.org/10.1103/PhysRevLett.67.181}
  {\bibfield  {journal} {\bibinfo  {journal} {Phys. Rev. Lett.}\ }\textbf
  {\bibinfo {volume} {67}},\ \bibinfo {pages} {181} (\bibinfo {year}
  {1991})}\BibitemShut {NoStop}%
\bibitem [{\citenamefont {Ejtemaee}\ and\ \citenamefont
  {Haljan}(2017)}]{ejtemaee2017jul}%
  \BibitemOpen
  \bibfield  {author} {\bibinfo {author} {\bibfnamefont {S.}~\bibnamefont
  {Ejtemaee}}\ and\ \bibinfo {author} {\bibfnamefont {P.~C.}\ \bibnamefont
  {Haljan}},\ }\bibfield  {title} {\bibinfo {title} {{{3D Sisyphus Cooling}} of
  {{Trapped Ions}}},\ }\href {https://doi.org/10.1103/PhysRevLett.119.043001}
  {\bibfield  {journal} {\bibinfo  {journal} {Phys. Rev. Lett.}\ }\textbf
  {\bibinfo {volume} {119}},\ \bibinfo {pages} {043001} (\bibinfo {year}
  {2017})}\BibitemShut {NoStop}%
\bibitem [{\citenamefont {Wang}\ \emph {et~al.}(2021)\citenamefont {Wang},
  \citenamefont {Li}, \citenamefont {Wu}, \citenamefont {Liu}, \citenamefont
  {Hu}, \citenamefont {Ma}, \citenamefont {Xiao},\ and\ \citenamefont
  {Jia}}]{wang2021}%
  \BibitemOpen
  \bibfield  {author} {\bibinfo {author} {\bibfnamefont {Y.}~\bibnamefont
  {Wang}}, \bibinfo {author} {\bibfnamefont {Y.}~\bibnamefont {Li}}, \bibinfo
  {author} {\bibfnamefont {J.}~\bibnamefont {Wu}}, \bibinfo {author}
  {\bibfnamefont {W.}~\bibnamefont {Liu}}, \bibinfo {author} {\bibfnamefont
  {J.}~\bibnamefont {Hu}}, \bibinfo {author} {\bibfnamefont {J.}~\bibnamefont
  {Ma}}, \bibinfo {author} {\bibfnamefont {L.}~\bibnamefont {Xiao}},\ and\
  \bibinfo {author} {\bibfnamefont {S.}~\bibnamefont {Jia}},\ }\bibfield
  {title} {\bibinfo {title} {Hybrid evaporative cooling of
  {\textsuperscript{133}}{{Cs}} atoms to {{Bose-Einstein}} condensation},\
  }\href {https://doi.org/10.1364/OE.419854} {\bibfield  {journal} {\bibinfo
  {journal} {Opt. Express}\ }\textbf {\bibinfo {volume} {29}},\ \bibinfo
  {pages} {13960} (\bibinfo {year} {2021})}\BibitemShut {NoStop}%
\bibitem [{\citenamefont {Zohar}\ \emph {et~al.}(2022)\citenamefont {Zohar},
  \citenamefont {Florshaim}, \citenamefont {Zilberman}, \citenamefont {Stern},\
  and\ \citenamefont {Sagi}}]{zohar2022}%
  \BibitemOpen
  \bibfield  {author} {\bibinfo {author} {\bibfnamefont {E.}~\bibnamefont
  {Zohar}}, \bibinfo {author} {\bibfnamefont {Y.}~\bibnamefont {Florshaim}},
  \bibinfo {author} {\bibfnamefont {O.}~\bibnamefont {Zilberman}}, \bibinfo
  {author} {\bibfnamefont {A.}~\bibnamefont {Stern}},\ and\ \bibinfo {author}
  {\bibfnamefont {Y.}~\bibnamefont {Sagi}},\ }\bibfield  {title} {\bibinfo
  {title} {Degenerate {{Raman}} sideband cooling of {{\textsuperscript{40}K}}
  atoms},\ }\href {https://doi.org/10.1103/PhysRevA.106.063111} {\bibfield
  {journal} {\bibinfo  {journal} {Phys. Rev. A}\ }\textbf {\bibinfo {volume}
  {106}},\ \bibinfo {pages} {063111} (\bibinfo {year} {2022})}\BibitemShut
  {NoStop}%
\bibitem [{\citenamefont {Debs}\ \emph {et~al.}(2011)\citenamefont {Debs},
  \citenamefont {Altin}, \citenamefont {Barter}, \citenamefont {D{\"o}ring},
  \citenamefont {Dennis}, \citenamefont {McDonald}, \citenamefont {Anderson},
  \citenamefont {Close},\ and\ \citenamefont {Robins}}]{debs2011sep}%
  \BibitemOpen
  \bibfield  {author} {\bibinfo {author} {\bibfnamefont {J.~E.}\ \bibnamefont
  {Debs}}, \bibinfo {author} {\bibfnamefont {P.~A.}\ \bibnamefont {Altin}},
  \bibinfo {author} {\bibfnamefont {T.~H.}\ \bibnamefont {Barter}}, \bibinfo
  {author} {\bibfnamefont {D.}~\bibnamefont {D{\"o}ring}}, \bibinfo {author}
  {\bibfnamefont {G.~R.}\ \bibnamefont {Dennis}}, \bibinfo {author}
  {\bibfnamefont {G.}~\bibnamefont {McDonald}}, \bibinfo {author}
  {\bibfnamefont {R.~P.}\ \bibnamefont {Anderson}}, \bibinfo {author}
  {\bibfnamefont {J.~D.}\ \bibnamefont {Close}},\ and\ \bibinfo {author}
  {\bibfnamefont {N.~P.}\ \bibnamefont {Robins}},\ }\bibfield  {title}
  {\bibinfo {title} {Cold-atom gravimetry with a {{Bose-Einstein}}
  condensate},\ }\href {https://doi.org/10.1103/PhysRevA.84.033610} {\bibfield
  {journal} {\bibinfo  {journal} {Phys. Rev. A}\ }\textbf {\bibinfo {volume}
  {84}},\ \bibinfo {pages} {033610} (\bibinfo {year} {2011})}\BibitemShut
  {NoStop}%
\bibitem [{\citenamefont {{Louchet-Chauvet}}\ \emph {et~al.}(2011)\citenamefont
  {{Louchet-Chauvet}}, \citenamefont {Farah}, \citenamefont {Bodart},
  \citenamefont {Clairon}, \citenamefont {Landragin}, \citenamefont {Merlet},\
  and\ \citenamefont {Pereira Dos~Santos}}]{louchet-chauvet2011}%
  \BibitemOpen
  \bibfield  {author} {\bibinfo {author} {\bibfnamefont {A.}~\bibnamefont
  {{Louchet-Chauvet}}}, \bibinfo {author} {\bibfnamefont {T.}~\bibnamefont
  {Farah}}, \bibinfo {author} {\bibfnamefont {Q.}~\bibnamefont {Bodart}},
  \bibinfo {author} {\bibfnamefont {A.}~\bibnamefont {Clairon}}, \bibinfo
  {author} {\bibfnamefont {A.}~\bibnamefont {Landragin}}, \bibinfo {author}
  {\bibfnamefont {S.}~\bibnamefont {Merlet}},\ and\ \bibinfo {author}
  {\bibfnamefont {F.}~\bibnamefont {Pereira Dos~Santos}},\ }\bibfield  {title}
  {\bibinfo {title} {The influence of transverse motion within an atomic
  gravimeter},\ }\href {https://doi.org/10.1088/1367-2630/13/6/065025}
  {\bibfield  {journal} {\bibinfo  {journal} {New J. Phys.}\ }\textbf {\bibinfo
  {volume} {13}},\ \bibinfo {pages} {065025} (\bibinfo {year}
  {2011})}\BibitemShut {NoStop}%
\bibitem [{\citenamefont {Anderson}\ \emph {et~al.}(1995)\citenamefont
  {Anderson}, \citenamefont {Ensher}, \citenamefont {Matthews}, \citenamefont
  {Wieman},\ and\ \citenamefont {Cornell}}]{anderson1995jul}%
  \BibitemOpen
  \bibfield  {author} {\bibinfo {author} {\bibfnamefont {M.~H.}\ \bibnamefont
  {Anderson}}, \bibinfo {author} {\bibfnamefont {J.~R.}\ \bibnamefont
  {Ensher}}, \bibinfo {author} {\bibfnamefont {M.~R.}\ \bibnamefont
  {Matthews}}, \bibinfo {author} {\bibfnamefont {C.~E.}\ \bibnamefont
  {Wieman}},\ and\ \bibinfo {author} {\bibfnamefont {E.~A.}\ \bibnamefont
  {Cornell}},\ }\bibfield  {title} {\bibinfo {title} {Observation of
  {{Bose-Einstein Condensation}} in a {{Dilute Atomic Vapor}}},\ }\href
  {https://doi.org/10.1126/science.269.5221.198} {\bibfield  {journal}
  {\bibinfo  {journal} {Science}\ }\textbf {\bibinfo {volume} {269}},\ \bibinfo
  {pages} {198} (\bibinfo {year} {1995})}\BibitemShut {NoStop}%
\bibitem [{\citenamefont {{van der Stam}}\ \emph {et~al.}(2007)\citenamefont
  {{van der Stam}}, \citenamefont {{van Ooijen}}, \citenamefont {Meppelink},
  \citenamefont {Vogels},\ and\ \citenamefont {{van der
  Straten}}}]{vanderstam2007jan}%
  \BibitemOpen
  \bibfield  {author} {\bibinfo {author} {\bibfnamefont {K.~M.~R.}\
  \bibnamefont {{van der Stam}}}, \bibinfo {author} {\bibfnamefont {E.~D.}\
  \bibnamefont {{van Ooijen}}}, \bibinfo {author} {\bibfnamefont
  {R.}~\bibnamefont {Meppelink}}, \bibinfo {author} {\bibfnamefont {J.~M.}\
  \bibnamefont {Vogels}},\ and\ \bibinfo {author} {\bibfnamefont
  {P.}~\bibnamefont {{van der Straten}}},\ }\bibfield  {title} {\bibinfo
  {title} {Large atom number {{Bose-Einstein}} condensate of sodium},\
  }\bibfield  {journal} {\bibinfo  {journal} {Rev. Sci. Instrum.}\ }\textbf
  {\bibinfo {volume} {78}},\ \href {https://doi.org/10.1063/1.2424439}
  {10.1063/1.2424439} (\bibinfo {year} {2007})\BibitemShut {NoStop}%
\bibitem [{\citenamefont {Xu}\ \emph {et~al.}(2024)\citenamefont {Xu},
  \citenamefont {{\v S}umarac}, \citenamefont {Qiu}, \citenamefont {Peters},
  \citenamefont {Cant{\'u}}, \citenamefont {Li}, \citenamefont {Menssen},
  \citenamefont {Lukin}, \citenamefont {Colombo},\ and\ \citenamefont
  {Vuleti{\'c}}}]{xu2024jun}%
  \BibitemOpen
  \bibfield  {author} {\bibinfo {author} {\bibfnamefont {W.}~\bibnamefont
  {Xu}}, \bibinfo {author} {\bibfnamefont {T.}~\bibnamefont {{\v S}umarac}},
  \bibinfo {author} {\bibfnamefont {E.~H.}\ \bibnamefont {Qiu}}, \bibinfo
  {author} {\bibfnamefont {M.~L.}\ \bibnamefont {Peters}}, \bibinfo {author}
  {\bibfnamefont {S.~H.}\ \bibnamefont {Cant{\'u}}}, \bibinfo {author}
  {\bibfnamefont {Z.}~\bibnamefont {Li}}, \bibinfo {author} {\bibfnamefont
  {A.}~\bibnamefont {Menssen}}, \bibinfo {author} {\bibfnamefont {M.~D.}\
  \bibnamefont {Lukin}}, \bibinfo {author} {\bibfnamefont {S.}~\bibnamefont
  {Colombo}},\ and\ \bibinfo {author} {\bibfnamefont {V.}~\bibnamefont
  {Vuleti{\'c}}},\ }\bibfield  {title} {\bibinfo {title} {Bose-{{Einstein
  Condensation}} by {{Polarization Gradient Laser Cooling}}},\ }\href
  {https://doi.org/10.1103/PhysRevLett.132.233401} {\bibfield  {journal}
  {\bibinfo  {journal} {Phys. Rev. Lett.}\ }\textbf {\bibinfo {volume} {132}},\
  \bibinfo {pages} {233401} (\bibinfo {year} {2024})}\BibitemShut {NoStop}%
\bibitem [{\citenamefont {Wei}\ and\ \citenamefont {Kuhn}(2018)}]{wei2018jun}%
  \BibitemOpen
  \bibfield  {author} {\bibinfo {author} {\bibfnamefont {C.}~\bibnamefont
  {Wei}}\ and\ \bibinfo {author} {\bibfnamefont {C.~C.~N.}\ \bibnamefont
  {Kuhn}},\ }\bibfield  {title} {\bibinfo {title} {Laser cooling of rubidium
  atoms in a {{2D}} optical lattice},\ }\href
  {https://doi.org/10.1080/09500340.2018.1429684} {\bibfield  {journal}
  {\bibinfo  {journal} {J. Mod. Opt.}\ }\textbf {\bibinfo {volume} {65}},\
  \bibinfo {pages} {1226} (\bibinfo {year} {2018})}\BibitemShut {NoStop}%
\bibitem [{\citenamefont {Wigley}\ \emph {et~al.}(2016)\citenamefont {Wigley},
  \citenamefont {Everitt}, \citenamefont {{van den Hengel}}, \citenamefont
  {Bastian}, \citenamefont {Sooriyabandara}, \citenamefont {McDonald},
  \citenamefont {Hardman}, \citenamefont {Quinlivan}, \citenamefont {Manju},
  \citenamefont {Kuhn}, \citenamefont {Petersen}, \citenamefont {Luiten},
  \citenamefont {Hope}, \citenamefont {Robins},\ and\ \citenamefont
  {Hush}}]{wigley2016}%
  \BibitemOpen
  \bibfield  {author} {\bibinfo {author} {\bibfnamefont {P.~B.}\ \bibnamefont
  {Wigley}}, \bibinfo {author} {\bibfnamefont {P.~J.}\ \bibnamefont {Everitt}},
  \bibinfo {author} {\bibfnamefont {A.}~\bibnamefont {{van den Hengel}}},
  \bibinfo {author} {\bibfnamefont {J.~W.}\ \bibnamefont {Bastian}}, \bibinfo
  {author} {\bibfnamefont {M.~A.}\ \bibnamefont {Sooriyabandara}}, \bibinfo
  {author} {\bibfnamefont {G.~D.}\ \bibnamefont {McDonald}}, \bibinfo {author}
  {\bibfnamefont {K.~S.}\ \bibnamefont {Hardman}}, \bibinfo {author}
  {\bibfnamefont {C.~D.}\ \bibnamefont {Quinlivan}}, \bibinfo {author}
  {\bibfnamefont {P.}~\bibnamefont {Manju}}, \bibinfo {author} {\bibfnamefont
  {C.~C.~N.}\ \bibnamefont {Kuhn}}, \bibinfo {author} {\bibfnamefont {I.~R.}\
  \bibnamefont {Petersen}}, \bibinfo {author} {\bibfnamefont {A.~N.}\
  \bibnamefont {Luiten}}, \bibinfo {author} {\bibfnamefont {J.~J.}\
  \bibnamefont {Hope}}, \bibinfo {author} {\bibfnamefont {N.~P.}\ \bibnamefont
  {Robins}},\ and\ \bibinfo {author} {\bibfnamefont {M.~R.}\ \bibnamefont
  {Hush}},\ }\bibfield  {title} {\bibinfo {title} {Fast machine-learning online
  optimization of ultra-cold-atom experiments},\ }\href
  {https://doi.org/10.1038/srep25890} {\bibfield  {journal} {\bibinfo
  {journal} {Sci. Rep.}\ }\textbf {\bibinfo {volume} {6}},\ \bibinfo {pages}
  {25890} (\bibinfo {year} {2016})}\BibitemShut {NoStop}%
\bibitem [{\citenamefont {Carleo}\ and\ \citenamefont
  {Troyer}(2017)}]{carleo2017}%
  \BibitemOpen
  \bibfield  {author} {\bibinfo {author} {\bibfnamefont {G.}~\bibnamefont
  {Carleo}}\ and\ \bibinfo {author} {\bibfnamefont {M.}~\bibnamefont
  {Troyer}},\ }\bibfield  {title} {\bibinfo {title} {Solving the quantum
  many-body problem with artificial neural networks},\ }\href
  {https://doi.org/10.1126/science.aag2302} {\bibfield  {journal} {\bibinfo
  {journal} {Science}\ }\textbf {\bibinfo {volume} {355}},\ \bibinfo {pages}
  {602} (\bibinfo {year} {2017})}\BibitemShut {NoStop}%
\bibitem [{\citenamefont {Torlai}\ \emph {et~al.}(2018)\citenamefont {Torlai},
  \citenamefont {Mazzola}, \citenamefont {Carrasquilla}, \citenamefont
  {Troyer}, \citenamefont {Melko},\ and\ \citenamefont {Carleo}}]{torlai2018}%
  \BibitemOpen
  \bibfield  {author} {\bibinfo {author} {\bibfnamefont {G.}~\bibnamefont
  {Torlai}}, \bibinfo {author} {\bibfnamefont {G.}~\bibnamefont {Mazzola}},
  \bibinfo {author} {\bibfnamefont {J.}~\bibnamefont {Carrasquilla}}, \bibinfo
  {author} {\bibfnamefont {M.}~\bibnamefont {Troyer}}, \bibinfo {author}
  {\bibfnamefont {R.}~\bibnamefont {Melko}},\ and\ \bibinfo {author}
  {\bibfnamefont {G.}~\bibnamefont {Carleo}},\ }\bibfield  {title} {\bibinfo
  {title} {Neural-network quantum state tomography},\ }\href
  {https://doi.org/10.1038/s41567-018-0048-5} {\bibfield  {journal} {\bibinfo
  {journal} {Nat. Phys.}\ }\textbf {\bibinfo {volume} {14}},\ \bibinfo {pages}
  {447} (\bibinfo {year} {2018})}\BibitemShut {NoStop}%
\bibitem [{\citenamefont {Rohringer}\ \emph {et~al.}(2008)\citenamefont
  {Rohringer}, \citenamefont {B{\"u}cker}, \citenamefont {Manz}, \citenamefont
  {Betz}, \citenamefont {Koller}, \citenamefont {G{\"o}bel}, \citenamefont
  {Perrin}, \citenamefont {Schmiedmayer},\ and\ \citenamefont
  {Schumm}}]{rohringer2008}%
  \BibitemOpen
  \bibfield  {author} {\bibinfo {author} {\bibfnamefont {W.}~\bibnamefont
  {Rohringer}}, \bibinfo {author} {\bibfnamefont {R.}~\bibnamefont
  {B{\"u}cker}}, \bibinfo {author} {\bibfnamefont {S.}~\bibnamefont {Manz}},
  \bibinfo {author} {\bibfnamefont {T.}~\bibnamefont {Betz}}, \bibinfo {author}
  {\bibfnamefont {{\relax Ch}.}~\bibnamefont {Koller}}, \bibinfo {author}
  {\bibfnamefont {M.}~\bibnamefont {G{\"o}bel}}, \bibinfo {author}
  {\bibfnamefont {A.}~\bibnamefont {Perrin}}, \bibinfo {author} {\bibfnamefont
  {J.}~\bibnamefont {Schmiedmayer}},\ and\ \bibinfo {author} {\bibfnamefont
  {T.}~\bibnamefont {Schumm}},\ }\bibfield  {title} {\bibinfo {title}
  {Stochastic optimization of a cold atom experiment using a genetic
  algorithm},\ }\href {https://doi.org/10.1063/1.3058756} {\bibfield  {journal}
  {\bibinfo  {journal} {Appl. Phys. Lett.}\ }\textbf {\bibinfo {volume} {93}},\
  \bibinfo {pages} {264101} (\bibinfo {year} {2008})}\BibitemShut {NoStop}%
\bibitem [{\citenamefont {Palittapongarnpim}\ \emph {et~al.}(2017)\citenamefont
  {Palittapongarnpim}, \citenamefont {Wittek}, \citenamefont {Zahedinejad},
  \citenamefont {Vedaie},\ and\ \citenamefont
  {Sanders}}]{palittapongarnpim2017}%
  \BibitemOpen
  \bibfield  {author} {\bibinfo {author} {\bibfnamefont {P.}~\bibnamefont
  {Palittapongarnpim}}, \bibinfo {author} {\bibfnamefont {P.}~\bibnamefont
  {Wittek}}, \bibinfo {author} {\bibfnamefont {E.}~\bibnamefont {Zahedinejad}},
  \bibinfo {author} {\bibfnamefont {S.}~\bibnamefont {Vedaie}},\ and\ \bibinfo
  {author} {\bibfnamefont {B.~C.}\ \bibnamefont {Sanders}},\ }\bibfield
  {title} {\bibinfo {title} {Learning in quantum control: High-dimensional
  global optimization for noisy quantum dynamics},\ }\href
  {https://doi.org/10.1016/j.neucom.2016.12.087} {\bibfield  {journal}
  {\bibinfo  {journal} {Neurocomputing}\ }\textbf {\bibinfo {volume} {268}},\
  \bibinfo {pages} {116} (\bibinfo {year} {2017})}\BibitemShut {NoStop}%
\bibitem [{\citenamefont {Che}\ \emph {et~al.}(2022)\citenamefont {Che},
  \citenamefont {Li}, \citenamefont {Fang}, \citenamefont {Chen},\ and\
  \citenamefont {Qin}}]{che2022jul_075211}%
  \BibitemOpen
  \bibfield  {author} {\bibinfo {author} {\bibfnamefont {H.}~\bibnamefont
  {Che}}, \bibinfo {author} {\bibfnamefont {A.}~\bibnamefont {Li}}, \bibinfo
  {author} {\bibfnamefont {J.}~\bibnamefont {Fang}}, \bibinfo {author}
  {\bibfnamefont {X.}~\bibnamefont {Chen}},\ and\ \bibinfo {author}
  {\bibfnamefont {F.-J.}\ \bibnamefont {Qin}},\ }\bibfield  {title} {\bibinfo
  {title} {Interference fringe fitting of atom gravimeter based on fitness
  particle swarm optimization},\ }\href {https://doi.org/10.1063/5.0096967}
  {\bibfield  {journal} {\bibinfo  {journal} {AIP Advances}\ }\textbf {\bibinfo
  {volume} {12}},\ \bibinfo {pages} {075211} (\bibinfo {year}
  {2022})}\BibitemShut {NoStop}%
\bibitem [{\citenamefont {Liu}\ \emph {et~al.}(2021)\citenamefont {Liu},
  \citenamefont {Xie}, \citenamefont {Li}, \citenamefont {Liang}, \citenamefont
  {Li}, \citenamefont {Chen}, \citenamefont {Fang}, \citenamefont {Qu},
  \citenamefont {Liu}, \citenamefont {Wang},\ and\ \citenamefont
  {L{\"u}}}]{liu2021}%
  \BibitemOpen
  \bibfield  {author} {\bibinfo {author} {\bibfnamefont {Q.}~\bibnamefont
  {Liu}}, \bibinfo {author} {\bibfnamefont {Y.}~\bibnamefont {Xie}}, \bibinfo
  {author} {\bibfnamefont {L.}~\bibnamefont {Li}}, \bibinfo {author}
  {\bibfnamefont {A.}~\bibnamefont {Liang}}, \bibinfo {author} {\bibfnamefont
  {W.}~\bibnamefont {Li}}, \bibinfo {author} {\bibfnamefont {H.}~\bibnamefont
  {Chen}}, \bibinfo {author} {\bibfnamefont {S.}~\bibnamefont {Fang}}, \bibinfo
  {author} {\bibfnamefont {Q.}~\bibnamefont {Qu}}, \bibinfo {author}
  {\bibfnamefont {L.}~\bibnamefont {Liu}}, \bibinfo {author} {\bibfnamefont
  {B.}~\bibnamefont {Wang}},\ and\ \bibinfo {author} {\bibfnamefont
  {D.}~\bibnamefont {L{\"u}}},\ }\bibfield  {title} {\bibinfo {title}
  {Multiparameter autonomous optimization system for ultracold atomic
  experiments based on artificial neural network},\ }\href
  {https://doi.org/10.3788/CJL202148.2412001} {\bibfield  {journal} {\bibinfo
  {journal} {Chin. J. Lasers}\ }\textbf {\bibinfo {volume} {48}},\ \bibinfo
  {pages} {2412001} (\bibinfo {year} {2021})}\BibitemShut {NoStop}%
\bibitem [{\citenamefont {Tranter}\ \emph {et~al.}(2018)\citenamefont
  {Tranter}, \citenamefont {Slatyer}, \citenamefont {Hush}, \citenamefont
  {Leung}, \citenamefont {Everett}, \citenamefont {Paul}, \citenamefont
  {{Vernaz-Gris}}, \citenamefont {Lam}, \citenamefont {Buchler},\ and\
  \citenamefont {Campbell}}]{tranter2018}%
  \BibitemOpen
  \bibfield  {author} {\bibinfo {author} {\bibfnamefont {A.~D.}\ \bibnamefont
  {Tranter}}, \bibinfo {author} {\bibfnamefont {H.~J.}\ \bibnamefont
  {Slatyer}}, \bibinfo {author} {\bibfnamefont {M.~R.}\ \bibnamefont {Hush}},
  \bibinfo {author} {\bibfnamefont {A.~C.}\ \bibnamefont {Leung}}, \bibinfo
  {author} {\bibfnamefont {J.~L.}\ \bibnamefont {Everett}}, \bibinfo {author}
  {\bibfnamefont {K.~V.}\ \bibnamefont {Paul}}, \bibinfo {author}
  {\bibfnamefont {P.}~\bibnamefont {{Vernaz-Gris}}}, \bibinfo {author}
  {\bibfnamefont {P.~K.}\ \bibnamefont {Lam}}, \bibinfo {author} {\bibfnamefont
  {B.~C.}\ \bibnamefont {Buchler}},\ and\ \bibinfo {author} {\bibfnamefont
  {G.~T.}\ \bibnamefont {Campbell}},\ }\bibfield  {title} {\bibinfo {title}
  {Multiparameter optimisation of a magneto-optical trap using deep learning},\
  }\href {https://doi.org/10.1038/s41467-018-06847-1} {\bibfield  {journal}
  {\bibinfo  {journal} {Nat. Commun.}\ }\textbf {\bibinfo {volume} {9}},\
  \bibinfo {pages} {4360} (\bibinfo {year} {2018})}\BibitemShut {NoStop}%
\bibitem [{\citenamefont {Vendeiro}\ \emph {et~al.}(2022)\citenamefont
  {Vendeiro}, \citenamefont {Ramette}, \citenamefont {Rudelis}, \citenamefont
  {Chong}, \citenamefont {Sinclair}, \citenamefont {Stewart}, \citenamefont
  {Urvoy},\ and\ \citenamefont {Vuleti{\'c}}}]{vendeiro2022dec}%
  \BibitemOpen
  \bibfield  {author} {\bibinfo {author} {\bibfnamefont {Z.}~\bibnamefont
  {Vendeiro}}, \bibinfo {author} {\bibfnamefont {J.}~\bibnamefont {Ramette}},
  \bibinfo {author} {\bibfnamefont {A.}~\bibnamefont {Rudelis}}, \bibinfo
  {author} {\bibfnamefont {M.}~\bibnamefont {Chong}}, \bibinfo {author}
  {\bibfnamefont {J.}~\bibnamefont {Sinclair}}, \bibinfo {author}
  {\bibfnamefont {L.}~\bibnamefont {Stewart}}, \bibinfo {author} {\bibfnamefont
  {A.}~\bibnamefont {Urvoy}},\ and\ \bibinfo {author} {\bibfnamefont
  {V.}~\bibnamefont {Vuleti{\'c}}},\ }\bibfield  {title} {\bibinfo {title}
  {Machine-learning-accelerated {{Bose-Einstein}} condensation},\ }\href
  {https://doi.org/10.1103/PhysRevResearch.4.043216} {\bibfield  {journal}
  {\bibinfo  {journal} {Phys. Rev. Res.}\ }\textbf {\bibinfo {volume} {4}},\
  \bibinfo {pages} {043216} (\bibinfo {year} {2022})}\BibitemShut {NoStop}%
\bibitem [{\citenamefont {LeDesma}\ \emph {et~al.}(2024)\citenamefont
  {LeDesma}, \citenamefont {Mehling}, \citenamefont {Shao}, \citenamefont
  {Wilson}, \citenamefont {Axelrad}, \citenamefont {Nicotra}, \citenamefont
  {Anderson},\ and\ \citenamefont {Holland}}]{ledesma2023}%
  \BibitemOpen
  \bibfield  {author} {\bibinfo {author} {\bibfnamefont {C.}~\bibnamefont
  {LeDesma}}, \bibinfo {author} {\bibfnamefont {K.}~\bibnamefont {Mehling}},
  \bibinfo {author} {\bibfnamefont {J.}~\bibnamefont {Shao}}, \bibinfo {author}
  {\bibfnamefont {J.~D.}\ \bibnamefont {Wilson}}, \bibinfo {author}
  {\bibfnamefont {P.}~\bibnamefont {Axelrad}}, \bibinfo {author} {\bibfnamefont
  {M.}~\bibnamefont {Nicotra}}, \bibinfo {author} {\bibfnamefont {D.~Z.}\
  \bibnamefont {Anderson}},\ and\ \bibinfo {author} {\bibfnamefont
  {M.}~\bibnamefont {Holland}},\ }\bibfield  {title} {\bibinfo {title}
  {Demonstration of a programmable optical lattice atom interferometer},\
  }\href {https://doi.org/10.1103/PhysRevResearch.6.043120} {\bibfield
  {journal} {\bibinfo  {journal} {Phys. Rev. Res.}\ }\textbf {\bibinfo {volume}
  {6}},\ \bibinfo {pages} {043120} (\bibinfo {year} {2024})}\BibitemShut
  {NoStop}%
\bibitem [{\citenamefont {Reinschmidt}\ \emph {et~al.}(2024)\citenamefont
  {Reinschmidt}, \citenamefont {Fort{\'a}gh}, \citenamefont {G{\"u}nther},\
  and\ \citenamefont {Volchkov}}]{reinschmidt2024oct}%
  \BibitemOpen
  \bibfield  {author} {\bibinfo {author} {\bibfnamefont {M.}~\bibnamefont
  {Reinschmidt}}, \bibinfo {author} {\bibfnamefont {J.}~\bibnamefont
  {Fort{\'a}gh}}, \bibinfo {author} {\bibfnamefont {A.}~\bibnamefont
  {G{\"u}nther}},\ and\ \bibinfo {author} {\bibfnamefont {V.~V.}\ \bibnamefont
  {Volchkov}},\ }\bibfield  {title} {\bibinfo {title} {Reinforcement learning
  in cold atom experiments},\ }\href
  {https://doi.org/10.1038/s41467-024-52775-8} {\bibfield  {journal} {\bibinfo
  {journal} {Nat. Commun.}\ }\textbf {\bibinfo {volume} {15}},\ \bibinfo
  {pages} {8532} (\bibinfo {year} {2024})}\BibitemShut {NoStop}%
\bibitem [{\citenamefont {Barker}\ \emph {et~al.}(2020)\citenamefont {Barker},
  \citenamefont {Style}, \citenamefont {Luksch}, \citenamefont {Sunami},
  \citenamefont {Garrick}, \citenamefont {Hill}, \citenamefont {Foot},\ and\
  \citenamefont {Bentine}}]{barker2020mar}%
  \BibitemOpen
  \bibfield  {author} {\bibinfo {author} {\bibfnamefont {A.~J.}\ \bibnamefont
  {Barker}}, \bibinfo {author} {\bibfnamefont {H.}~\bibnamefont {Style}},
  \bibinfo {author} {\bibfnamefont {K.}~\bibnamefont {Luksch}}, \bibinfo
  {author} {\bibfnamefont {S.}~\bibnamefont {Sunami}}, \bibinfo {author}
  {\bibfnamefont {D.}~\bibnamefont {Garrick}}, \bibinfo {author} {\bibfnamefont
  {F.}~\bibnamefont {Hill}}, \bibinfo {author} {\bibfnamefont {C.~J.}\
  \bibnamefont {Foot}},\ and\ \bibinfo {author} {\bibfnamefont
  {E.}~\bibnamefont {Bentine}},\ }\bibfield  {title} {\bibinfo {title}
  {Applying machine learning optimization methods to the production of a
  quantum gas},\ }\href {https://doi.org/10.1088/2632-2153/ab6432} {\bibfield
  {journal} {\bibinfo  {journal} {Mach. Learn.: Sci. Technol.}\ }\textbf
  {\bibinfo {volume} {1}},\ \bibinfo {pages} {015007} (\bibinfo {year}
  {2020})}\BibitemShut {NoStop}%
\bibitem [{\citenamefont {Xu}\ \emph {et~al.}(2021)\citenamefont {Xu},
  \citenamefont {Kaebert}, \citenamefont {Stepanova}, \citenamefont {Poll},
  \citenamefont {Siercke},\ and\ \citenamefont {Ospelkaus}}]{xu2021}%
  \BibitemOpen
  \bibfield  {author} {\bibinfo {author} {\bibfnamefont {S.}~\bibnamefont
  {Xu}}, \bibinfo {author} {\bibfnamefont {P.}~\bibnamefont {Kaebert}},
  \bibinfo {author} {\bibfnamefont {M.}~\bibnamefont {Stepanova}}, \bibinfo
  {author} {\bibfnamefont {T.}~\bibnamefont {Poll}}, \bibinfo {author}
  {\bibfnamefont {M.}~\bibnamefont {Siercke}},\ and\ \bibinfo {author}
  {\bibfnamefont {S.}~\bibnamefont {Ospelkaus}},\ }\bibfield  {title} {\bibinfo
  {title} {Maximizing the capture velocity of molecular magneto-optical traps
  with {{Bayesian}} optimization},\ }\href
  {https://doi.org/10.1088/1367-2630/ac06e6} {\bibfield  {journal} {\bibinfo
  {journal} {New J. Phys.}\ }\textbf {\bibinfo {volume} {23}},\ \bibinfo
  {pages} {063062} (\bibinfo {year} {2021})}\BibitemShut {NoStop}%
\bibitem [{\citenamefont {Liang}\ \emph {et~al.}(2024)\citenamefont {Liang},
  \citenamefont {Gao}, \citenamefont {Liu}, \citenamefont {Wang}, \citenamefont
  {Yan}, \citenamefont {Yang}, \citenamefont {Zhu},\ and\ \citenamefont
  {Ma}}]{liang2024nov}%
  \BibitemOpen
  \bibfield  {author} {\bibinfo {author} {\bibfnamefont {C.}~\bibnamefont
  {Liang}}, \bibinfo {author} {\bibfnamefont {S.}~\bibnamefont {Gao}}, \bibinfo
  {author} {\bibfnamefont {J.}~\bibnamefont {Liu}}, \bibinfo {author}
  {\bibfnamefont {G.}~\bibnamefont {Wang}}, \bibinfo {author} {\bibfnamefont
  {S.}~\bibnamefont {Yan}}, \bibinfo {author} {\bibfnamefont {J.}~\bibnamefont
  {Yang}}, \bibinfo {author} {\bibfnamefont {L.}~\bibnamefont {Zhu}},\ and\
  \bibinfo {author} {\bibfnamefont {X.}~\bibnamefont {Ma}},\ }\bibfield
  {title} {\bibinfo {title} {Multi-parameter optimization of polarization
  gradient cooling for {\textsuperscript{87}}{{Rb}} atoms based on
  reinforcement learning},\ }\href {https://doi.org/10.1364/OE.537239}
  {\bibfield  {journal} {\bibinfo  {journal} {Opt. Express}\ }\textbf {\bibinfo
  {volume} {32}},\ \bibinfo {pages} {40364} (\bibinfo {year}
  {2024})}\BibitemShut {NoStop}%
\bibitem [{\citenamefont {Anton}\ \emph {et~al.}(2024)\citenamefont {Anton},
  \citenamefont {Henderson}, \citenamefont {Da~Ros}, \citenamefont {Sekulic},
  \citenamefont {Burger}, \citenamefont {Schneider},\ and\ \citenamefont
  {Krutzik}}]{anton2024jun}%
  \BibitemOpen
  \bibfield  {author} {\bibinfo {author} {\bibfnamefont {O.}~\bibnamefont
  {Anton}}, \bibinfo {author} {\bibfnamefont {V.~A.}\ \bibnamefont
  {Henderson}}, \bibinfo {author} {\bibfnamefont {E.}~\bibnamefont {Da~Ros}},
  \bibinfo {author} {\bibfnamefont {I.}~\bibnamefont {Sekulic}}, \bibinfo
  {author} {\bibfnamefont {S.}~\bibnamefont {Burger}}, \bibinfo {author}
  {\bibfnamefont {P.-I.}\ \bibnamefont {Schneider}},\ and\ \bibinfo {author}
  {\bibfnamefont {M.}~\bibnamefont {Krutzik}},\ }\bibfield  {title} {\bibinfo
  {title} {Review and experimental benchmarking of machine learning algorithms
  for efficient optimization of cold atom experiments},\ }\href
  {https://doi.org/10.1088/2632-2153/ad3cb6} {\bibfield  {journal} {\bibinfo
  {journal} {Mach. Learn.: Sci. Technol.}\ }\textbf {\bibinfo {volume} {5}},\
  \bibinfo {pages} {025022} (\bibinfo {year} {2024})}\BibitemShut {NoStop}%
\bibitem [{\citenamefont {Yao}\ \emph {et~al.}(1999)\citenamefont {Yao},
  \citenamefont {Liu},\ and\ \citenamefont {Lin}}]{yao1999jul}%
  \BibitemOpen
  \bibfield  {author} {\bibinfo {author} {\bibfnamefont {X.}~\bibnamefont
  {Yao}}, \bibinfo {author} {\bibfnamefont {Y.}~\bibnamefont {Liu}},\ and\
  \bibinfo {author} {\bibfnamefont {G.}~\bibnamefont {Lin}},\ }\bibfield
  {title} {\bibinfo {title} {Evolutionary programming made faster},\ }\href
  {https://doi.org/10.1109/4235.771163} {\bibfield  {journal} {\bibinfo
  {journal} {IEEE Trans. Evolut. Comput.}\ }\textbf {\bibinfo {volume} {3}},\
  \bibinfo {pages} {82} (\bibinfo {year} {1999})}\BibitemShut {NoStop}%
\bibitem [{\citenamefont {Xin}\ and\ \citenamefont {Yuan}(2024)}]{yuan2024jan}%
  \BibitemOpen
  \bibfield  {author} {\bibinfo {author} {\bibfnamefont {W.}~\bibnamefont
  {Xin}}\ and\ \bibinfo {author} {\bibfnamefont {S.}~\bibnamefont {Yuan}},\
  }\bibfield  {title} {\bibinfo {title} {Research progress of cold atomic
  temperature measurement methods},\ }\href
  {http://yqyb.etmchina.com/yqyben/article/abstract/J2311468} {\bibfield
  {journal} {\bibinfo  {journal} {Chin. J. Sci. Instrum.}\ }\textbf {\bibinfo
  {volume} {44}},\ \bibinfo {pages} {70} (\bibinfo {year} {2024})}\BibitemShut
  {NoStop}%
\bibitem [{\citenamefont {Petelski}(2005)}]{petelski2005}%
  \BibitemOpen
  \bibfield  {author} {\bibinfo {author} {\bibfnamefont {T.~G.}\ \bibnamefont
  {Petelski}},\ }\emph {\bibinfo {title} {Atom interferometers for precision
  gravity measurements}},\ \href@noop {} {\bibinfo {type} {Phd dissertation}},\
  \bibinfo  {school} {Pierre and Marie Curie University}, \bibinfo {address}
  {Paris} (\bibinfo {year} {2005})\BibitemShut {NoStop}%
\end{thebibliography}%

\appendix	
\section{Laser System\label{app:laser}}
\begin{figure}
	\includegraphics[width=\columnwidth]{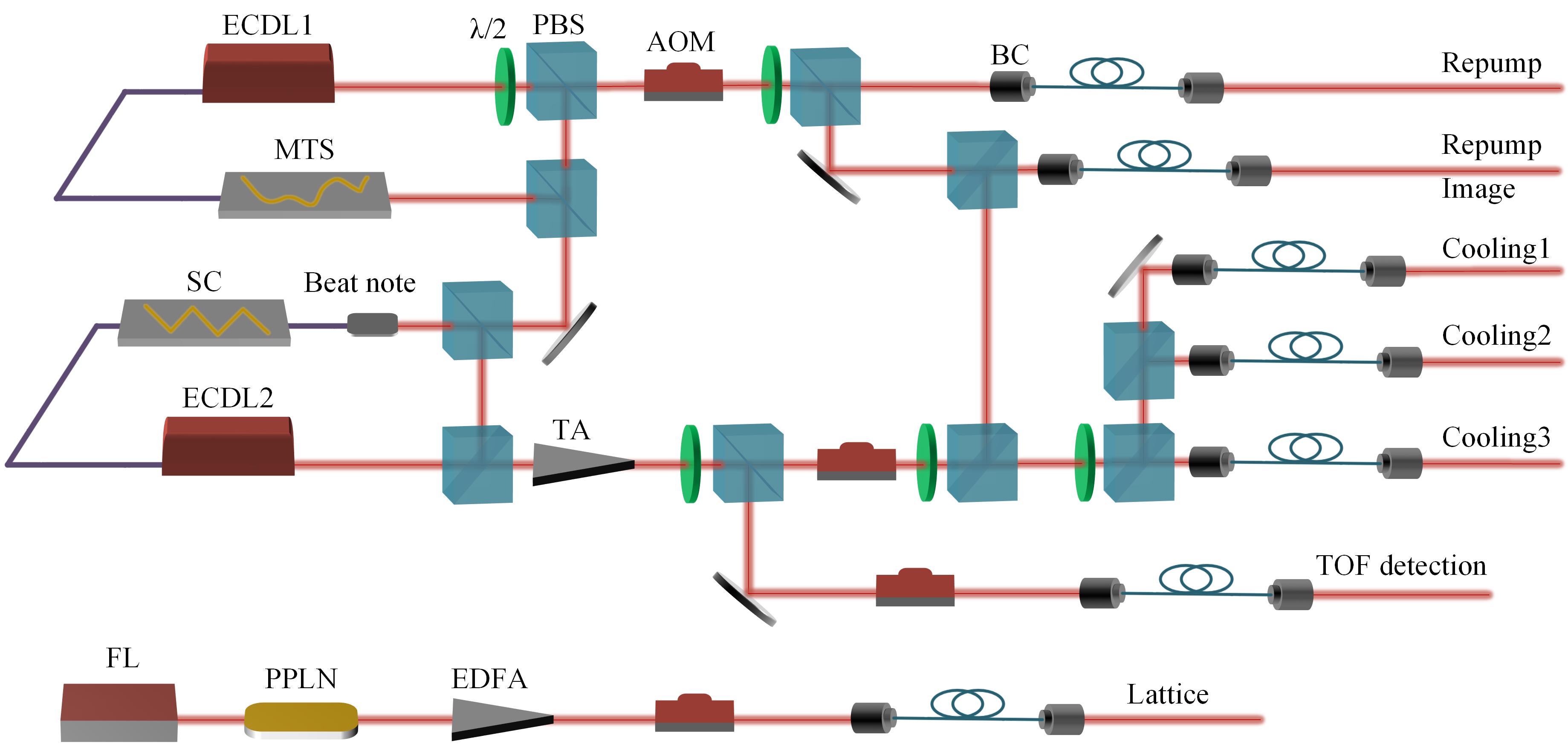}
	\caption{Schematic diagram of the laser system. ECDL: External Cavity Diode Laser, FL: Fiber Laser, PBS: Polarizing Beam Splitter, BC: Beam Collimator, AOM: Acousto-Optic Modulator, MTS: Modulation Transfer Spectroscopy Locking Module, SC: Servo Controller, PPLN: Periodically Poled Lithium Niobate, EDFA: Erbium-Doped Fiber Amplifier.}
	\label{fig:ls}
\end{figure}

\begin{figure}
	\includegraphics[width=0.6\columnwidth]{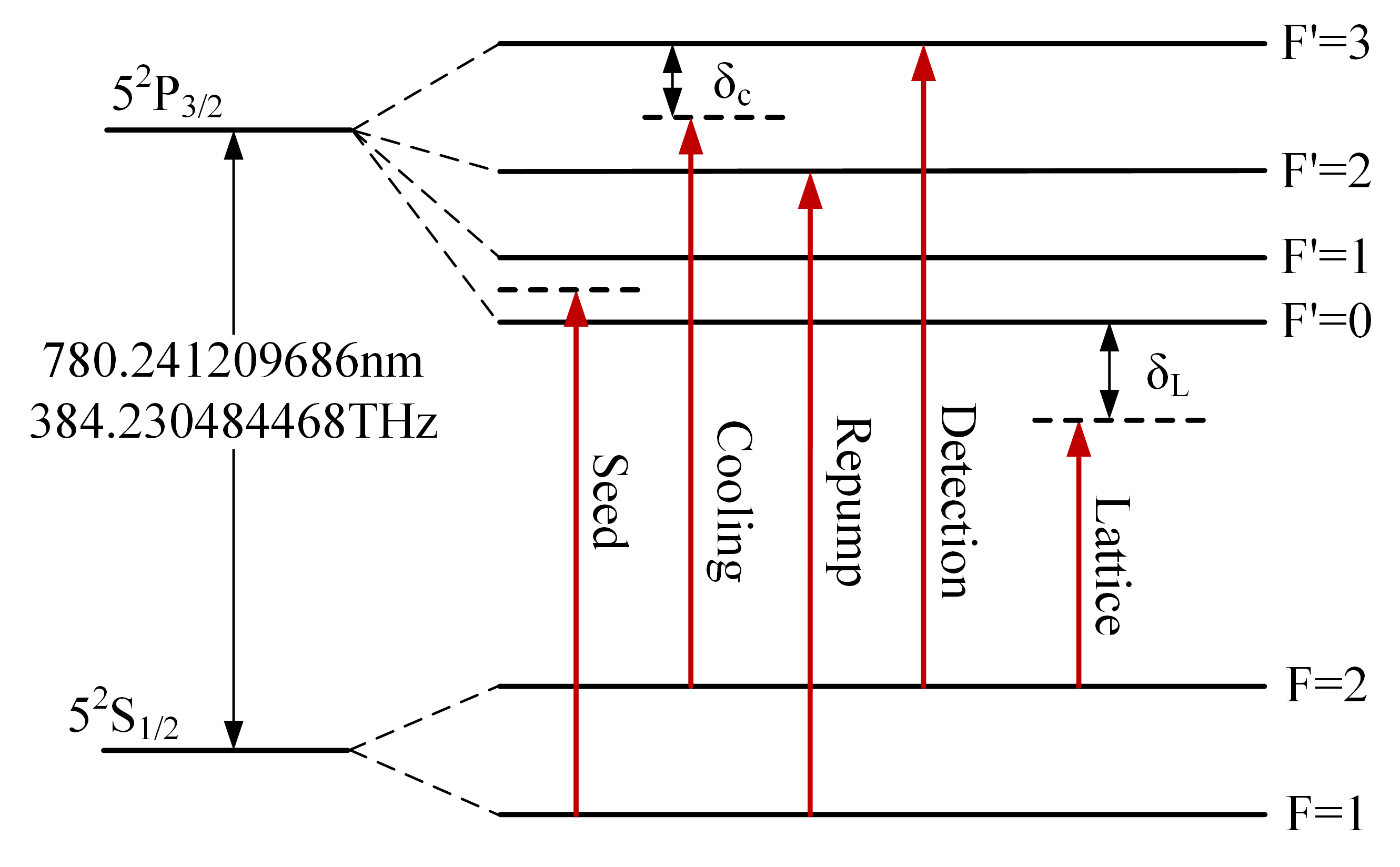}
	\caption{The energy levels of $^{87}$Rb D$_2$ line and laser frequency allocation. $\delta_{\rm c}$ represents the cooling light detuning, while $\delta_{\rm L}$ represents the lattice light detuning.}
	\label{fig:lf}
\end{figure}

The schematic diagram of the laser system is shown in FIG.~\ref{fig:ls}. The light source is mainly provided by three lasers, including two External Cavity Diode Lasers (ECDLs) and one fiber laser. ECDL1, serving as the seed light, is frequency-locked to the$|F=1 \rangle \rightarrow |F'=0 \rangle$  and $|F=1 \rangle \rightarrow |F'=1 \rangle$  crossover transition on the $^{87}$Rb D$_2$ line using the modulation transfer spectrum (MTS) locking module. After frequency shifting by Acousto-Optic Modulator (AOM) to $|F=1 \rangle \rightarrow |F'=2 \rangle$ , it is used as the repumping light (2.2 mW). ECDL2 is frequency-locked to $|F=2 \rangle \rightarrow |F'=3 \rangle$  using the beat-note signal with ECDL1. After frequency shifting by AOM, it is used as the cooling light ($\sim$14 mW per beam), imaging light ($\sim$14 mW per beam), and TOF detection light ($\sim$72 $\upmu$W). The diameters of the repumping light, imaging light, and cooling light are $\sim$26 mm, while the TOF detection light forms a sheet of light at 10 mm $\times$ 1 mm after being blocked by an aperture. The 1560 nm laser from a fiber laser, once amplified by an Erbium-Doped Fiber Amplifier (EDFA), is frequency-doubled through Periodically Poled Lithium Niobate (PPLN) to 780 nm, and frequency shifted by AOM (primarily for switching and amplitude modulation), used as the lattice light ($\sim$19.5 mW) after collimation with a diameter of about 1.8mm. The allocation of each laser frequency is shown in FIG.~\ref{fig:lf}, with the cooling light detuning $\delta_{\rm c}$ at $\sim$15 MHz in the MOT, 50-200 MHz in the PGC, and the lattice light detuning $\delta_{\rm L}$ at $\sim$67 GHz.

\begin{figure*}[ht!]
	\includegraphics[width=2\columnwidth]{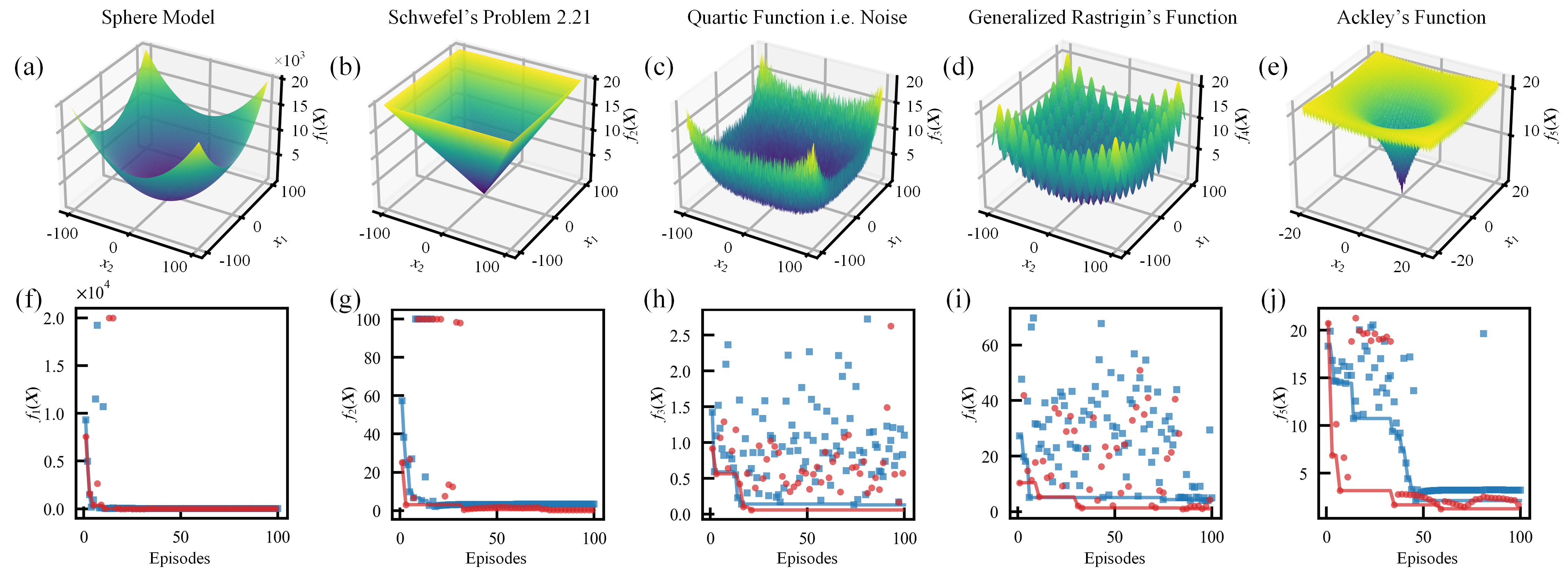}
	\caption{MHCS-BO performance on 5 typical test functions. The 3-dimension graphs (a-e) and optimization processes (f-j) of 5 typical test functions (Tab.~\ref{tab:tf}), where the red dots and blue squares represent the function value in MHCS-BO and BO respectively, while the solid lines represent corresponding historical optimum.}
	\label{fig:av1}
\end{figure*}

\section{Traditional BO\label{app:TBO}}
Initially, 10 random samples are taken in the parameter space to conduct experiment cycle successively. After calculating the corresponding $O(\mathbf{X}$), the initial observed dataset is obtained, which is then used to perform an initial fitting for the Gaussian process. To avoid overfitting, enhance generalization capability, and accelerate convergence, a variable-length adaptive radial basis function kernel is employed as the kernel function, with an initial length of 0.01, varying from 0.001 to 5. The adaptive kernel function can automatically adjust its shape based on the data characteristics, thus flexibly adapting to changes in different data distributions and features. Based on this model, the average value $\mu$ and standard deviation $\delta$ at $\mathbf{X}_{\rm s}$, a point sampled in parameter space, can be calculated by fitting the signal, allowing the calculation of the pre-designed objective function.
\begin{eqnarray}
	\mu = K(\mathbf{X}_{\rm s}, \mathbf{X}_{\rm o})[ K(\mathbf{X}_{\rm o}, \mathbf{X}_{\rm o})+\sigma_{\rm o}^2\mathbf{I}]^{-1}\mathbf{Y}_{\rm o}
	\label{eq:mu}.
\end{eqnarray}
\begin{equation}
	\begin{aligned}
		\sigma &= \mathbf{K}(\mathbf{X}_{\rm s}, \mathbf{X}_{\rm s})- \\
		& \mathbf{K}(\mathbf{X}_{\rm s}, \mathbf{X}_{\rm o}) \left [ \mathbf{K}(\mathbf{X}_{\rm o}, \mathbf{X}_{\rm o})+\sigma_{\rm o}^2\mathbf{I} \right ]^{-1} \mathbf{K}(\mathbf{X}_{\rm o}, \mathbf{X}_{\rm s})
	\end{aligned}
	\label{eq:mu}
\end{equation}
Where $\mathbf{X}_{\rm o}=\{\mathbf{X}_{\rm 1}, \mathbf{X}_{\rm 2}, \dots, \mathbf{X}_{\rm N}\}$, $\mathbf{Y}_{\rm o}=\{y_{\rm 1}, y_{\rm 2}, \dots, y_{\rm N}\}^\top$ represent the parameter vector and objective function corresponding to each data point. $\sigma_{\rm o}$ denotes the observation noise. $\mathbf{K}(\mathbf{X}, \mathbf{Y})_{i, j} = k(\mathbf{X}_i, \mathbf{Y}_j)$ is the covariance matrix, where $k(\mathbf{X}_i, \mathbf{Y}_j)$ is the radial basis function kernel between point $\mathbf{X}_i$ and $\mathbf{Y}_i$. The corresponding EI value for $\mathbf{X}_{\rm s}$ can be computed as follows:
\begin{equation}
	y_{\rm EI}=\left\{\begin{matrix}
		( \mu - y^* )\Phi(z)+\sigma\phi(z)  & \sigma \neq 0\\
		0  & \sigma = 0
	\end{matrix}\right. \\
	\label{eq:ei}
\end{equation}
Where $z=( \mu - y^* )/\sigma$, and $y^*$ is the best objective function in observed dataset, $\Phi(\cdot)$ and $\phi(\cdot)$ represent the probability density function (PDF) and cumulative distribution function (CDF) of the standard normal distribution, respectively. The L-BFGS-B optimization algorithm is utilized to find the maximum value. To avoid getting trapped in local optimum, 30 parameters are randomly selected in the search space as the initial values for L-BFGS-B. Then, the 30 individual optimizations are performed separately. The maximum value among all results is taken as the final optimization result, with the corresponding parameters {$\mathbf{X}_{\rm EI}$ being the optimal parameters predicted by EI, which is then applied in running one experiment cycle to calculate the objective function $O_{\rm EI}$. The scaled data point $\{\mathbf{X}_{\rm EI}$, $O_{\rm EI}\}$ is then added to the observed dataset. 
\begin{table*}[ht!] %
	\caption{\label{tab:tf} Details of 5 test functions. $n$ represents the dimension, $f_{\rm min}$ is the minimum value of the function.}
	\begin{ruledtabular}
		\begin{tabular}{cccccc} %
			Number & Name & Expression & $n$ &$x_i$ & $f_{\rm min}$\\ %
			\hline
			1 & Sphere Model                                   & $f_1(\mathbf{X}) = \sum_{1}^{n}x_i^2$                     & 2   & [-100,100]  & 0  \\ 
			2 &  Schwefel's Problem 2.21                 & $f_2(\mathbf{X}) = {\rm max}_i\{|x_i|, 1\leq i \leq n\}$ & 2   & [-100,100]  & 0  \\ 
			3 &  Quartic Function i.e. Noise             & $f_3(\mathbf{X}) = \sum_{1}^{n}ix_i^4+{\rm random}[0,1)$ &2   & [-1,1]  & 0  \\ 
			4 &  Generalized Rastrigin’s Function    & $f_4(\mathbf{X}) = \sum_{1}^{n}[x_i^2-10{\rm cos}(2\pi x_i)+10]$ & 2   & [-5,5]  & 0  \\ 
			5 &  Ackley’s Function                           & $f_5(\mathbf{X}) = -20{\rm exp}\left(-0.2\sqrt{(1/n)\sum_{1}^{n}x_i^2}\right)-
			{\rm exp}\left((1/n)\sum_{1}^{n}{\rm cos}(2\pi x_i)\right)+20+{\rm e}$ & 2   & [-20,20]  & 0  \\ 
		\end{tabular}
	\end{ruledtabular}
\end{table*}

\section{Validating MHCS-BO algorithm in 5 test functions\label{app:mhcs_test}}
The details of the 5 test functions employed in this study are outlined in Tab.~\ref{tab:tf}. To clearly represent each function’s characteristics in a 3-dimensional graph as shown in FIG.~\ref{fig:av1}(a)-(e), the parameter dimension is set to 2. The optimization processes are illustrated in FIG.~\ref{fig:av1}(f)-(j). When comparing the performance of EI predictions between MHCS-BO and BO, datapoints obtained by MHCS are not displayed, which implies each figure FIG.~\ref{fig:av1}(f)-(j) contains 50 red dots and 100 blue squares. It can be seen that the historical optimum achieved by MHCS-BO surpass those from BO. Furthermore, as datapoints predicted by MHCS are not used to fit the Gaussian process in the current episode, there is a significant reduction in time consumption.

\section{Measurement of $T_{\rm a}$ with TOF Method\label{app:tof}}
The TOF signal detection system is shown in FIG.~\ref{fig:tof}. The resonant TOF light passes through a beam expander (LC1258-B, AC508-150-B, Thorlabs, total length 170 mm), resulting in a diameter of $\sim$44 mm, and forms a uniform sheet of light at 10 mm $\times$ 1 mm after passing through an aperture. When atoms fall from the center of the MOT to this detection region, they interact with the laser and emit fluorescence. The PD converts the collected atomic fluorescence into an electrical signal (TOF signal), which is amplified and then collected by a 16-bit data acquisition card. The distance from the lens to the atoms is $l$=85 mm, and the lens diameter is $d$=50.8 mm.

\begin{figure}
	\includegraphics[width=0.8\columnwidth]{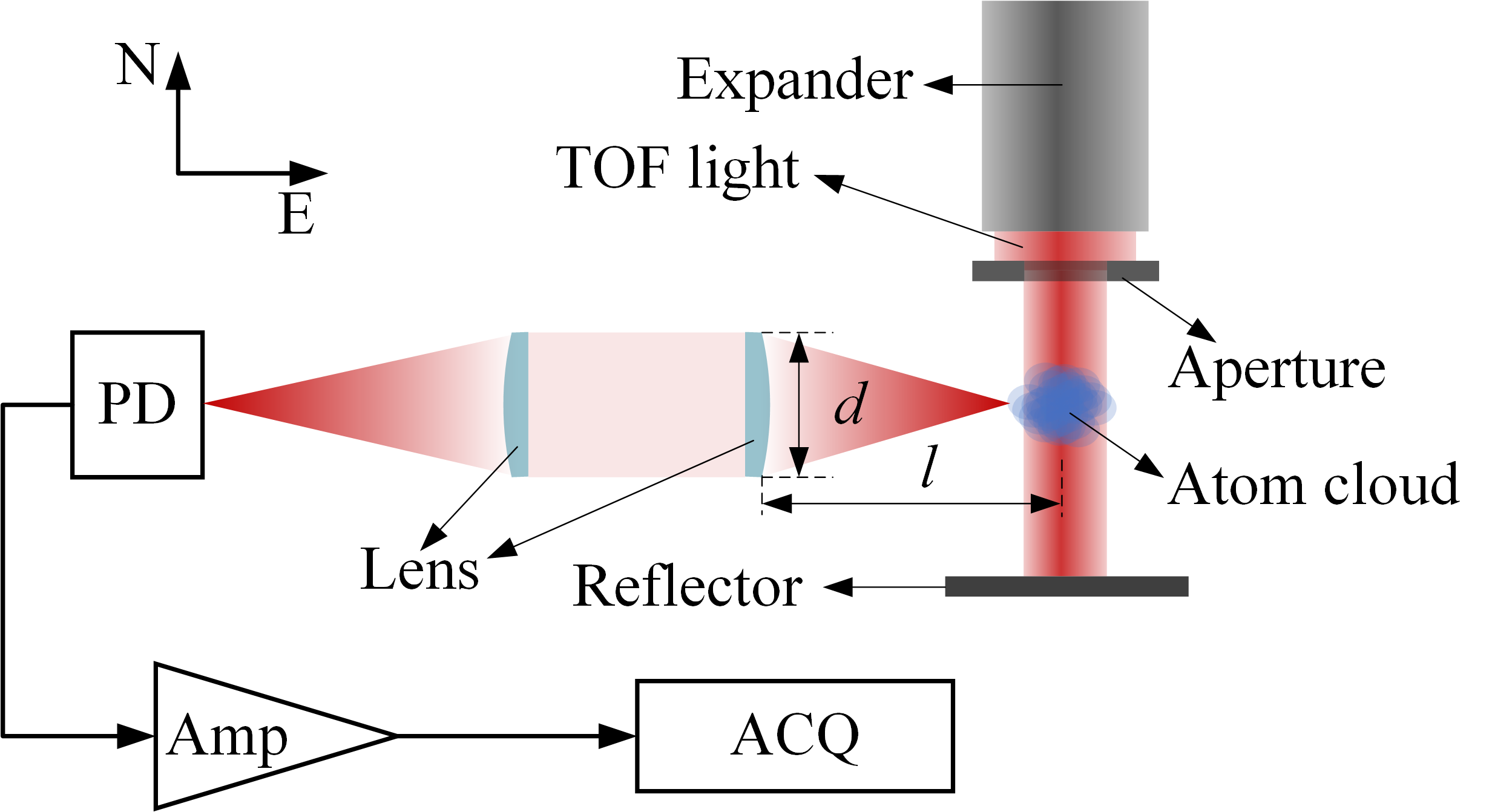}
	\caption{TOF signal detection system. PD: Photodetector, AMP: Amplifier, ACQ: Acquisition Card.}
	\label{fig:tof}
\end{figure}
The intensity of atomic fluorescence is proportional to the number of atoms. Given the small thickness of the detection region, it can be approximated that the time of each atom traversing the detection region is the same. Therefore, the final received TOF signal can be represented as
\begin{equation}
	I_{\rm TOF}(t) = KP(t)
	\label{eq:I_tof}
\end{equation}

Here, $P(t)$ represents the number of atoms that arrive at the detection region at time $t$, $K$ is the proportionality coefficient. Assuming the atomic cluster is relatively small initially, the initial vertical velocity of atoms in the cluster is $v_0$, and $v_0^2 \ll 2gh,$ then the time $t$ it takes for atoms to reach the detection region has the following linear relationship with $v_0$
\begin{equation}
	t=\left(\sqrt{2gh}-v_0\right)/g
	\label{eq:t}
\end{equation}
Using the above relationship, the acquired time-domain signal can be converted to the velocity domain, $I_{\rm TOF}(v_0)$, thus obtaining the velocity distribution of atoms in the atomic cluster (approximated as a Gaussian distribution). Through fitting, the full width at half maximum (FWHM) of the velocity distribution, denoted as $\Delta v_0$, can be obtained. The initial velocity of atoms before free diffusion satisfies the Maxwell-Boltzmann distribution, while each velocity component follows a normal distribution:
\begin{equation}
	f(v_0)=\left(\frac{m}{2\pi k_{\rm B}T}\right)^{1/2}{\rm exp}\left(-\frac{m}{2 k_{\rm B}T}v_0^2\right) 
	\label{eq:fvt}
\end{equation}
Here, $m$ and $T$ is the atom mass and temperature respectively, $k_{\rm B}$ is the Boltzmann constant. With the above equation, FWHM of the velocity distribution in a specific direction can be calculated as
\begin{equation}
	\Delta v_0=2\sqrt{\frac{2k_{\rm B}T}{m}{\rm ln}2}
	\label{eq:deltav}
\end{equation}
According to the above equation, the atom temperature $T$ can be obtained.

\section{Measurement of System Stability\label{app:longMeasure}}
\begin{figure}[!ht]
	\includegraphics[keepaspectratio=true]{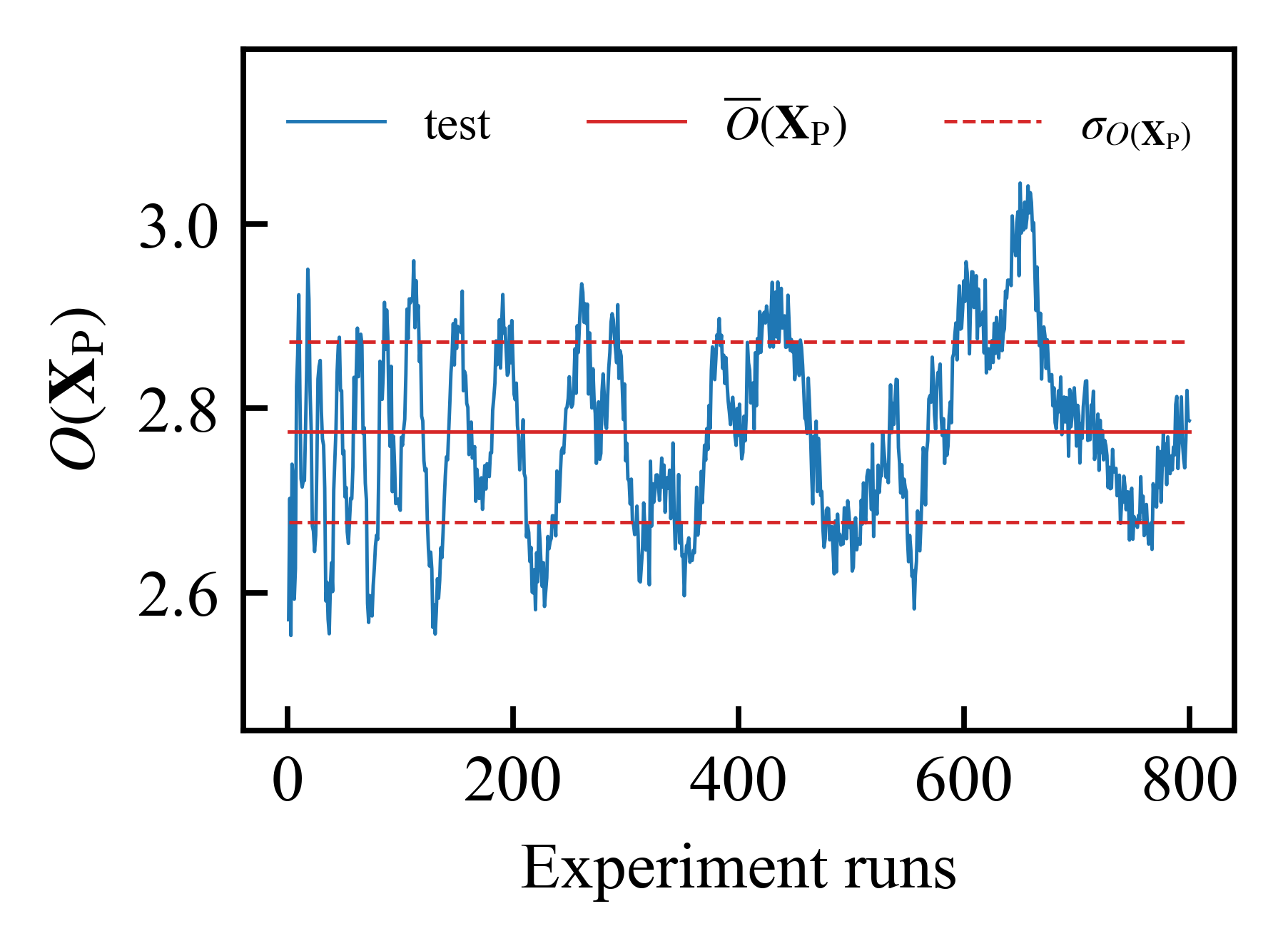}
	\caption{Long-term stability test of $O(\mathbf{X}_{\rm P})$. The red solid and dashed lines represent the position of mean value and one standard deviation respectively.}
	\label{fig:lt}
\end{figure}

To verify the long-term stability of the system, we set a typical PGC timing sequence and conduct 800 independent experiments to evaluate the fluctuation of $O(\mathbf{X}_{\rm P})$. The results are presented in FIG.~\ref{fig:lt}. Here, we utilized the coefficient of variation (CV) as a metric to evaluate the system’s stability, which is defined as the ratio of standard deviation to mean value. Based on the data depicted in FIG.~\ref{fig:lt}, the CV can be calculated as 0.035, exhibiting excellent long-term stability of the system. Furthermore, this metric can be employed to assess the volatility of the data shown in FIG.~\ref{fig:dp}(b). As illustrated, the $O(\mathbf{X})$ can take approximately 1.5 as a typical value. Multiplying this value by the CV results in a standard deviation of approximately 0.05, which aligns well with the error band.

\section{Influence of Temperature on Atom Interference\label{app:Interference}}
\begin{figure}[!h]
	\includegraphics[keepaspectratio=true]{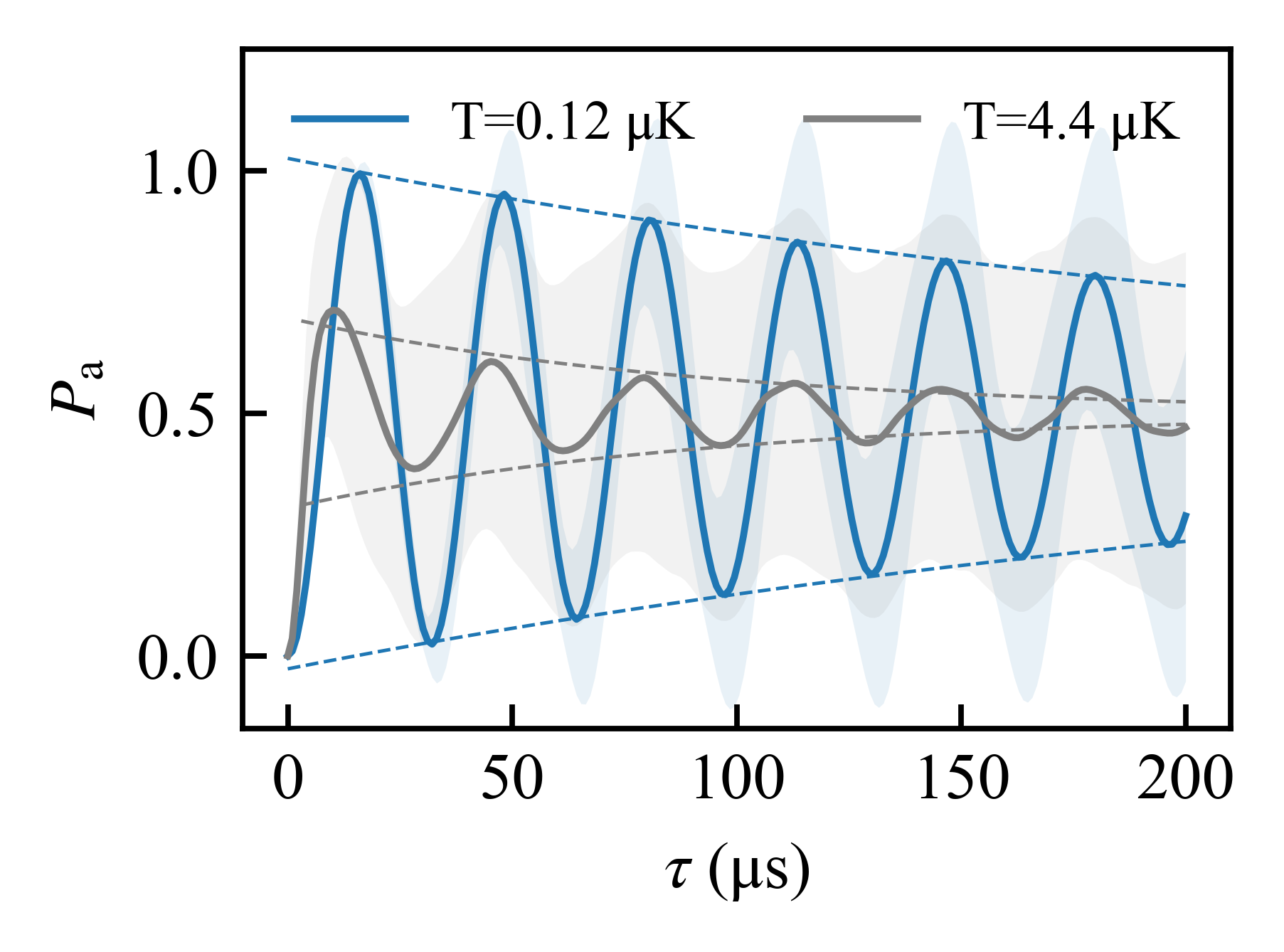}
	\caption{Rabi fringes under two different atom temperature. The blue and gray curves represent Rabi fringes at atom temperatures of 0.12 $\upmu$K and 0.44 $\upmu$K, respectively, with the shaded regions indicating one standard deviation. The dashed lines are the corresponding exponential envelope fits, employed to calculate coherence time. Here, we define coherence time as the interaction time at which the fringe contrast decays to 1/e of initial value.}
	\label{fig:rabi}
\end{figure}
In quantum precision measurement (e.g. quantum gravimeter), atom interference is typically achieved through two-photon Raman transitions, which can be equivalent to a two-level system. Assuming atoms are initially prepared in the excited state $|{\rm b}\rangle$, neglecting the initial phase of Raman light, AC-Stark shift, and shift of the two-photon Raman transition, the probability of atoms remaining in ground state $|{\rm a}\rangle$ after interacting with the optical field for time $\tau$ can be expressed as~\cite{petelski2005}
\begin{equation}
	P_{\rm a} = {\rm sin}^2\left(\frac{\sqrt{\Omega_{\rm eff}^2+\delta_{12}^2}}{2} \tau\right) 
	\label{eq:Pa}
\end{equation}
where $\Omega_{\rm eff}$ is the effective wavevector of the Raman light.

Setting the atom temperature as $T$, we perform $N$ independent velocity samplings $v_i$ ($i=1, 2, ..., N$) following the normal distribution specified in Eq.~\ref{eq:fvt}. Given the laser wavelength $\lambda$ and employing counter-propagating beams (a common technique in quantum precision measurement to enhance momentum transfer), the effective laser wavevector of Raman light becomes $k_{\rm eff}\approx 4\pi/\lambda$. The Doppler shift induced by atom velocity can then be expressed as $k_{\rm eff}v_i$, allowing Eq.~\ref{eq:Pa} to be rewritten as
\begin{equation}
	P_{\rm a} = {\rm sin}^2\left(\frac{\sqrt{\Omega_{\rm eff}^2+\left(\delta_{12}+ k_{\rm eff}v_i \right)^2}}{2} \tau\right) 
	\label{eq:pa}
\end{equation}
With $N=5000$, $\Omega_{\rm eff}=30$ kHz, $\delta_{12}=0$ and two typical atom temperature in this study, we simulate the Rabi fringes using Eq.~\ref{eq:pa}, as shown in FIG.~\ref{fig:rabi}. The results demonstrate that, when atom temperature is set at 0.12 $\upmu$K and 4.4 $\upmu$K, the fringe contrast and coherence time are 99.3$\%$, 71.2$\%$ and 289 $\upmu$s, 93 $\upmu$s respectively, demonstrating enhancements of approximately 139$\%$ and 311$\%$ correspondingly.

\end{document}